\documentclass{emulateapj}

\usepackage{amsmath}
\usepackage{amssymb}
\usepackage{graphicx}
\usepackage{color}

\newcommand{\lSect}[1]{{\label{sec:#1}}}
\newcommand{\lFig}[1]{{\label{fig:#1}}}
\newcommand{\lEq}[1]{{\label{eq:#1}}}
\newcommand{\lTab}[1]{{\label{tab:#1}}}
\newcommand{\Msun}{\ensuremath{\mathrm{M}_\odot}\,}

\newcommand{\Teff}{\ensuremath{T_{ef}}\,}
\newcommand{\Tcol}{\ensuremath{T_c}\,}

\newcommand{\kP}{\ensuremath{\kappa_P}\,}
\newcommand{\chiR}{\ensuremath{\chi_R}\,}
\newcommand{\Rth}{\ensuremath{R_c}\,}
\newcommand{\Rp}{\ensuremath{R_p}\,}

\newcommand{\pd}[2]{\frac{\partial#1}{\partial#2}}
\newcommand{\grad}[1]{\vec{\nabla} #1} 

\renewcommand{\div}[1]{\vec{\nabla} \cdot #1}

\newcommand{\vel}{\vec{v}}

\newcommand{\cm}{\ensuremath{\mathrm{cm}}}

\newcommand{\E}[1]{\ensuremath{ \times 10^{#1}}}	

\newcommand{\FIGFF}[2]{{\ref{fig:#2}{#1}}}

\newcommand{\FIG}[2]{{Fig.~\FIGFF{#1}{#2}}}
\newcommand{\Fig}[1]{{\FIG{}{#1}}}

\newcommand{\Sectff}[1]{{\ref{sec:#1}}}
\newcommand{\Sect}[1]{{\S~\Sectff{#1}}}

\newcommand{\Eqref}[1]{{\ref{eq:#1}}}
\newcommand{\Eqff}[1]{{(\Eqref{#1})}}

\newcommand{\Eq}[1]{{eq.~\Eqff{#1}}}

\newcommand{\Tab}[1]{{Table~\ref{tab:#1}}}

\newcommand{\red}{}

\begin{document}

\title{Very Low Energy Supernovae: Light Curves and Spectra of Shock Breakout}

\author{Elizabeth Lovegrove\altaffilmark{1},
  S. E. Woosley\altaffilmark{1}, and Weiqun Zhang\altaffilmark{2}}

\altaffiltext{1}{Department of Astronomy and Astrophysics, University
 of California, Santa Cruz, CA 95064; woosley@ucolick.org}
 
 \altaffiltext{2}{Lawrence Berkeley National Laboratory}
 
\begin{abstract}
The brief transient emitted as a shock wave erupts through the surface
of a presupernova star carries information about the stellar radius
and explosion energy. Here the CASTRO code, which treats radiation
transport using multigroup flux-limited diffusion, is used to simulate
the light curves and spectra of shock breakout in very low-energy
supernovae (VLE SNe), explosions in giant stars with final kinetic
energy much less than 10$^{51}$ erg. VLE SNe light curves, computed
here with the KEPLER code, are distinctively faint, red, and
long-lived, making them challenging to find with transient
surveys. The accompanying shock breakouts are brighter, though
briefer, and potentially easier to detect. Previous analytic work
provides general guidance, but numerical simulations are challenging
due to the range of conditions and lack of equilibration between color
and effective temperatures. We consider previous analytic work and
extend discussions of color temperature and opacity to the lower
energy range explored by these events. Since this is the first
application of the CASTRO code to shock breakout, test simulations of
normal energy shock breakout of SN1987A are carried out and compared
with the literature. A set of breakout light curves and spectra are
then calculated for VLE SNe with final kinetic energies in the range
$10^{47} - 10^{50}$ ergs for red supergiants with main sequence masses
15 \Msun and 25 \Msun. The importance of uncertainties in stellar
atmosphere model, opacity, and ambient medium is discussed, as are
observational prospects with current and forthcoming missions.
\end{abstract}

\section{Introduction}
\lSect{intro}

Supernova shock breakout occurs when the leading edge of the explosion
erupts through a star's surface. As the outgoing shock propagates
through the stellar envelope, radiation builds up behind an optically
thick leading edge. When the surrounding matter reaches a sufficiently
small optical depth, this radiation diffuses out in a relatively short
times to produce a bright flash. This flash is the second indication,
after the neutrino pulse, that a core-collapse supernova has
occurred. Because their properties are determined by the star's
structure in a thin layer near the surface, shock breakouts convey
unique information about the progenitor's surface gravity, radius,
composition, and explosion energy.

In a recent survey, \citet{Suk15} found that neutrino-powered
supernovae from 9 to 120 \Msun\! with central engines calibrated to
reproduce the known properties of the Crab supernova and SN 1987A,
fell into two categories: failures, which did not blow the star up at
all and made black holes, and robust explosions with energies above
10$^{50}$ erg. While the treatment of neutrino transport was
approximate, this behavior suggests a sensitivity to presupernova
properties and an explosion mechanism that is ``gated", i.e. it either
works well or not at all.  Achieving a neutrino-powered explosion of
e.g., $\sim10^{49}$ erg would require fine tuning. The same study also
noted, however, that the observed mass spectrum of black holes is
inconsistent with the frequent collapse of the entire star \citep[see
  also][]{Koc14}. Some uncertain phenomenon is needed to frequently
remove the envelope of failed supernovae, either during black hole
formation or before. Given that black holes masses are only determined
in binary systems, the envelope may have been frequently lost to a
close companion star, but there should also be many cases where
detached red supergiants, with the same helium core structure,
collapse to black holes.  Various possibilities for envelope ejection
in these cases have been discussed including the Nadyozhin-Lovegrove
effect \citep{unnova}, envelope ejection via acoustic energy input
\citep{waves1, waves2}, thermonuclear instability during silicon
burning \citep{Woo15} pulsational-pair instability supernovae
\citep{ppsn, assl}, and the outbursts of luminous blue variables
\citep{smithlbv}. These events have in common the ejection of
substantial envelope matter, by a shock wave in most cases, with
energy much less than 10$^{51}$ erg. The term ``very-low-energy
supernova" (VLE SN) is therefore defined in this work as a
shock-powered transient from a massive, evolved star with a final
kinetic energy significantly less than the $\sim$ 0.6 B expected from
core collapse (1 B = $10^{51}$ erg of final kinetic energy).



\citet{gerke15} have reviewed the current status of searches for
failed supernovae as well as the faint transients arising from the
Nadyozhin-Lovegrove effect. Four candidates were initially selected,
but followup observations ruled out three of these as they later
reappeared. The fourth candidate event satisfied the criteria for a
VLE SN and continued to be observed. Recently \citet{adams16} reported
that the source had dimmed significantly below the progenitor
luminosity. Modeling of possible dust effects compared to optical and
IR source data are consistent with the event indeed being terminal.
The transient's cool, dim properties cannot be explained by ordinary
dust. This event might therefore be the first observed example of the
Nadyozhin-Lovegrove effect and an excellent example of a real VLE SN
though further observations are needed to make sure the star has
indeed disappeared.

VLE SNe are faint and challenging to observe in the general supernova
population. Other quantities, like mass, radius, and opacity, being
equal, \citet{Pop93} and \citet{Kas09} predict that the luminosity of
a Type IIp supernova during its plateau stage scales as $E_{\rm
  exp}^{5/6}$. Thus a supernova with explosion energy, $E_{\rm exp}
\sim 10^{48}$ erg would have a luminosity 300 times fainter than a
typical Type IIp supernova with $E_{\rm exp} \sim 10^{51}$ erg, or
about 10$^{40}$ erg s$^{-1}$. The duration of the plateau, which goes
as $E_{\rm exp}^{-1/6}$ would be 4.6 times longer, or about 400 days.

Shock breakout on the other hand is briefer, brighter, and potentially
easier to detect by large field-of-view transient surveys. The
breakout from a 5\E{48} erg explosion would have a typical luminosity
of 8\E{42} erg s$^{-1}$ and last several hours. Unlike breakout in
common supernovae, the temperature is relatively low, about 1.15\E{5}
K, and a larger fraction of the energy would be emitted longward of
the Lyman limit. Observing the shock breakouts of VLE SNe can give
occurrence rates for this unusual sort of supernovae as well as
constrain the properties of the presupernova star and the explosion
energy. As we shall see, the properties of these low energy transients
are well defined, but a major uncertainty is the event rate.
 
Shock breakout has previously been simulated in models for SN1987A
\citep{eb92, t12}, Type Ib and Ic SNe \citep{t12}, Type IIp supernovae
\citep{tominagatypeII}, and pair-instability supernovae at cosmic
distances \citep{danppsn, whalenppsn}. Although analytic estimates
exist \citep{piro}, no numerical simulations have yet been carried out
for VLE SNe. In particular, the color temperature, which is distinct
from the effective temperature and essential to determining the
bolometric correction, has not been calculated because its
determination requires a multigroup treatment of radiation. In the
coming age of large-scale transient surveys accurate light curves and
spectra will be vital for mission planning and analysis. Without such
models, these transients might easily be confused with other phenomena
having similar time scales and luminosities - for instance, novae,
failed Type Ia SNe, tidal disruption events, or common envelope
ejection from binary mergers - and the light curve automated selection
criteria can inadvertently misclassify interesting transients or even
ignore them altogether.

The CASTRO code \citep{castro1} is a multi-dimensional compressible
hydrodynamics code that uses adaptive-mesh refinement. The code incorporates
multigroup flux-limited diffusion (MGFLD) transport of radiation
\citep{castro3}. Here, CASTRO is used with a constant mesh and only in
one dimension. The coordinates are Eulerian and spherical. The
diffusion approximation for radiation closes the radiation transport
equations by assuming a Fick's Law diffusion relation between
radiation energy density $E_r$ and flux $\vec{F}$, $\vec{F} =
\vec{\nabla}{E_r}$. This approximation is only appropriate in an
optically-thick regime, however. When the material becomes optically
thin, the diffusion treatment leads to superluminal velocities and
produces unphysical behavior. To avoid this, the equations can be
modified to incorporate a flux limiter, $\lambda$, that gives the
correct limiting behavior in both the diffusive and free-streaming
regimes and a stable transition in between \citep{fld81}. Propagation
velocities will not exceed the speed of light and the pressure tensor
will have the correct limiting behavior. This approximation allows a
stable modeling of the crucial transition between optically-thick and
optically-thin media that shapes the breakout flash.

Grey or two-temperature (``2T'') radiation transport models radiation
as a fluid with a separate temperature $T_{rad}$ that can vary from
the gas temperature, but assumes the radiation spectrum is always a
blackbody with temperature $T_{rad}$. CASTRO's multigroup capability,
unlike 2T transport, allows the radiation to have a non-thermalized
spectrum. In the case of shock breakout, the radiation spectrum is
expected to be a dilute blackbody, i.e. it has a blackbody form but
peaks at a different wavelength than would be predicted from the
radiation energy density. In this paper the temperature computed from
a simple $E_r = aT^4$ relation is called the ``effective temperature''
\Teff\!, while the Wien's law temperature corresponding to the emitted
spectrum's peak wavelength is the color temperature \Tcol\!. For
further details on MGFLD and its specific implementation in CASTRO,
see \citet{krumholz} (derivation) and \citet{castro2}
(implementation).

Opacity is considered in \Sect{opac_theory} and applied to color
temperature behavior in VLE SNe in \Sect{vle_tcol}. Results from our
SN1987A test problem are first given in \Sect{87a}. In \Sect{setup} we
describe the code setup and the starting models for the RSG runs. In
\Sect{lcs} we describe the methods for sampling and post-processing
the data. The red supergiant results are described in detail and
compared to analytic and numerical predictions in
\Sect{breakoutrsg}. Prospects for observing are considered in
\Sect{observing}.

\section{Opacity}
\lSect{opac_theory}

The properties of shock breakout are determined by the optically
thick-thin transition layer in the star's atmosphere and modeling it
requires careful attention to the opacity. Unfortunately, detailed
opacity calculations in the low density, hot regime of shock breakout
are not straightforward, especially when bound atoms play a
role. Existing studies of shock breakout have often assumed the
dominance of electron scattering opacities during shock breakout. This
is a reasonable assumption at standard supernova energies where
temperatures behind the shock will be in the range 1\E{5} - 1\E{7}
K. This assumption begins to break down, however, in the regime probed
by low energy events. For VLE SNe breakouts, we expect temperatures in
the range 1\E{4} - 5\E{5} K and densities in the range 1\E{-9} -
1\E{-12} g/cm$^3$. At these conditions different opacity processes
begin to play a role, and the tabulated opacities used by many
supernova and stellar evolution codes do not extend to the low-density
regime, requiring the code to do its own opacity calculations.

\citet{nakar10} define a parameter, $\eta$, that predicts whether a
shell can thermalize its trapped radiation. This parameter depends
strongly on the strength of local absorption. Therefore the absorption
opacity, quantified in CASTRO by the Planck mean \kP\!, must be
considered in detail to ensure accurate color temperature
calculations. This section discusses four major opacity processes,
their effects on both regular and low-energy shock breakout, and their
treatment in these models.

\subsection{Free-Free Processes (Compton scattering and bremsstrahlung)}
\lSect{freefree}

\subsubsection{Analytic Representation}

At the temperatures, densities, and metallicities explored in this
paper, a major contribution to scattering opacity comes from photons
colliding with free electrons (Compton scattering), which in these
low-energy models can be considered to be in the Thomson limit. This
process also contributes to absorption opacity, however, because the
collisions are not perfectly elastic and a small amount of energy is
exchanged between photon and electron. The energy exchanged per
collision is:
\begin{align}
\Delta E &= \frac{h\nu}{m_ec^2}(4kT_g - h\nu)
\end{align}
where {\red $T_g$ is the gas temperature}. If the photon energy $h\nu$
is less than the gas/electron energy $4kT_g$ then the photons will
gain energy (the inverse Compton process); otherwise they will lose
energy to the gas. As the radiation temperature drops, the average
photon energy $h\nu$ does as well, and each scattering exchanges less
energy. If the only absorption source considered is Compton
scattering, then as the breakout energy drops absorption will become
inefficient at equilibrating the radiation. The simulation will
therefore show the chromosphere retreating within the star to higher
temperatures and densities, increasing the \Tcol/\Teff ratio. This
will be discussed further in \Sect{vle_tcol}.

Bremsstrahlung emission occurs when an electron passes close to
another charged particle, generally an atomic nucleus, and the change
in its energy causes the emission of a photon. There is an associated
inverse bremsstrahlung process wherein a photon strikes an electron
moving near a nucleus and causes an increase in its energy. This
results in a transfer of energy from radiation to gas and therefore
functions as an absorptive opacity. Inverse bremsstrahlung in regions
with fully ionized hydrogen and helium follows a Kramer's Law opacity
and can be calculated analytically as:
\begin{align}
\lEq{brem_grey}
\kP_b &= C_b T^{-3.5}n_e n_i l^2
\end{align}
where $C_b$ is a numerical constant and $l$ is the ionization level of the nucleus.

While Compton scattering depends only on the electron number density
$n_e$, since it considers only photons scattering off electrons,
inverse bremsstrahlung depends on both $n_e$ and $n_i$, as both an
electron and an ion are involved in each interaction. For a completely
ionized atmosphere consisting of mostly hydrogen and helium, $n_i
\approx n_e$. This gives $\kP_b \propto n_i^2$, and in fully ionized
regions $\rho \approx n_i\mu$ where $\mu$ is the mean molecular
weight, resulting in a $\rho^2$ dependency. Inverse bremsstrahlung
therefore drops off more rapidly with $\rho$ than Compton scattering,
relevant since shock breakout takes place in a region of sharply
decreasing density. The steepness of this density profile is a
property of the progenitor star and therefore does not differ between
standard CCSNe and VLE SNe.

The gas temperature at the shock front, however, does vary
significantly between standard CCSNe and VLE SNe. Compton scattering
depends linearly on gas temperature, but inverse bremsstrahlung goes
as $\propto T^{-3.5}$. Thus a small drop in temperature causes the
inverse bremsstrahlung opacity to increase much faster than Compton
scattering, fast enough to overcome the suppression from the $\rho^2$
term, making inverse bremsstrahlung much more significant in models at
lower temperatures. In the VLE case temperatures during breakout are
on the order of 5\E{4} - 3\E{5} K, and in this regime inverse
bremsstrahlung plays a significant role and must be considered in
simulation. The different contributions of various opacity processes
at different energies are illustrated at low energy in \Fig{a1_opac}
and at a more average energy in \Fig{f1_opac}.

\subsubsection{Implementation}

Energy exchange via photon-electron collisions occurs at the same rate as electron scattering. The CASTRO opacity network used for breakout represents absorptive opacity from Compton scattering as a fraction of electron scattering opacity $\kappa_C = \epsilon \chi_e$, where $\chi_e$ is the Thomson electron scattering opacity and $\epsilon$ is some factor $< 1.0$. In our simulations $\epsilon$ is generally in the range 1\E{-3} - 1\E{-4}.

For inverse bremsstrahlung calculations the CASTRO opacity network implements \Eq{brem_grey}. Note that this approximation becomes invalid if hydrogen and helium are not fully ionized. When the gas temperature drops below approximately 4.5\E{4} K, helium and hydrogen begin to recombine, and this equation gives progressively more unreasonable answers as inverse bremsstrahlung no longer follows a Kramer's Law form. The opacity network therefore stops calculating bremsstrahlung opacity below this temperature.

\subsection{Bound-Free Processes (photoionization)}
\lSect{boundfree} 

Photoionization is a complex process. Because cross-sections for ionization depend strongly on the energy of the incoming photon, precise formulas for this opacity are heavily frequency-dependent. However, above 5000 K hydrogen may be assumed to be ionized, and above 4.5\E{4} K helium may also be assumed to be completely ionized. Therefore when $T > 4.5\E{4}$ K only metals will contribute to photoionization opacity. In a star of solar metallicity the fraction of the atmosphere that consists of metals is small, but their large cross-sections and low ionization energies make their opacities significant. In this regime the photoionization can be treated as a grey opacity with a Kramer's Law form.
\begin{align}
\lEq{phot_grey}
\kP_p&= 4.34\E{25}\,Z(1.0+X_H)\rho T^{-3.5}
\end{align}
where $X_H$ is the hydrogen mass fraction and $Z$ the metal mass fraction (not the proton number). However, this raises the question of how to treat photoionization when $T < 4.5\E{4}$ K. Here the Kramer's Law opacity breaks down and truly accurate calculations are quite time-intensive. An approximate grey opacity may be found by doing a number of frequency-dependent calculations and taking their Planck mean, but this is again quite time-intensive. It is therefore worth considering whether photoionization opacities below the helium ionization limit are relevant; in these models the opacity network ceases to calculate bound-free opacities below this temperature.

\begin{figure}
\begin{center}
\includegraphics[scale=0.4, clip=true, trim = 25 0 0 0]{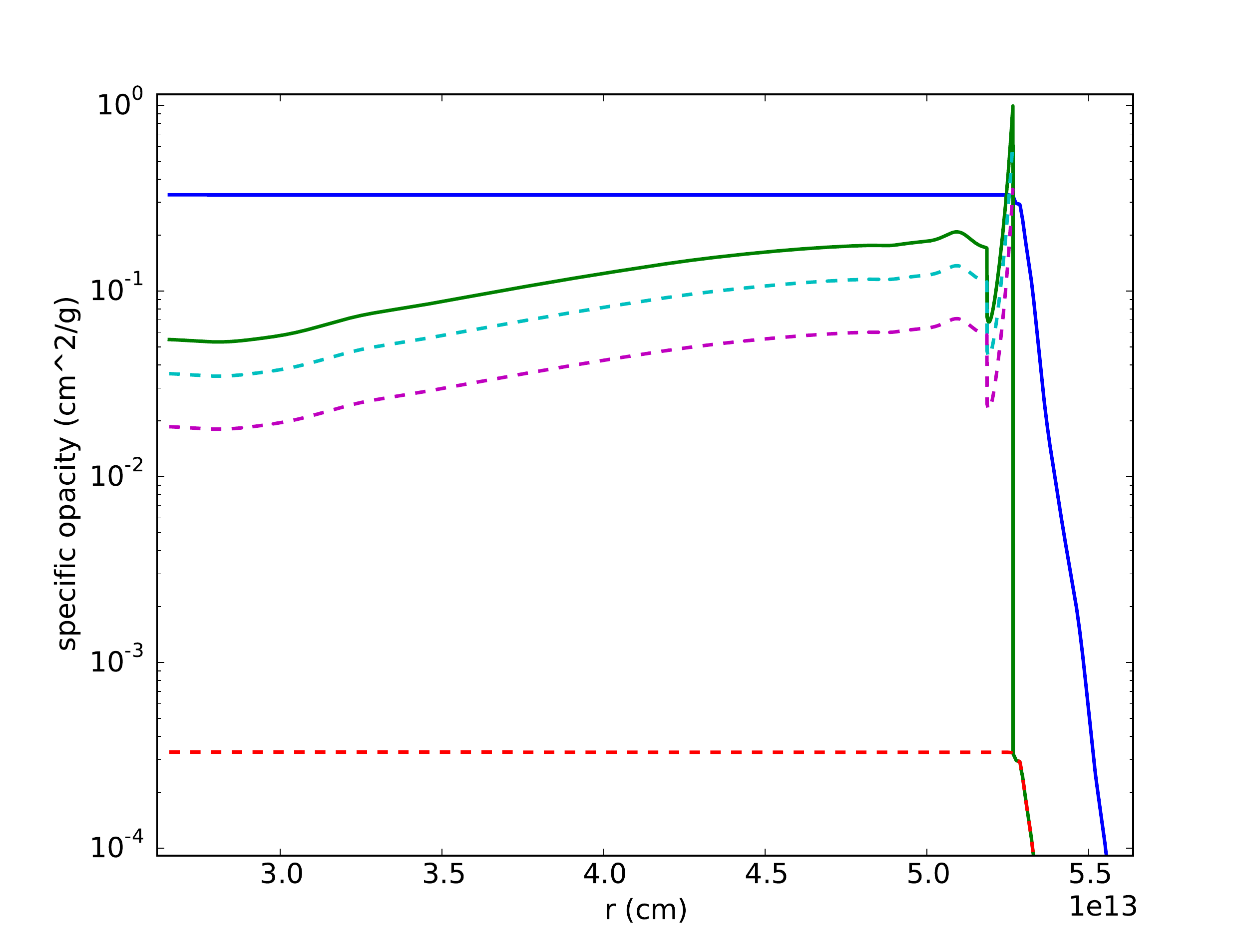}
\caption{ \lFig{a1_opac} Opacity profile for a 15 \Msun RSG near breakout with 1.54\E{48} erg of final kinetic energy (model B15 in \Sect{progenitor}), showing specific opacity $\kappa$. These opacities are calculated using the analytic formulas outlined in \Sect{opac_theory}. Solid lines show total opacity (blue) and total absorptive opacity (green). Dotted lines show absorptive opacity contributions from Compton scattering (red), inverse bremsstrahlung (cyan), and bound-free (magenta). Absorptive opacity is relatively strong in this low-energy model, particularly near the stellar surface.}
  \end{center}
\end{figure} 

\begin{figure}
\begin{center}
\includegraphics[scale=0.4, clip=true, trim = 25 0 0 0]{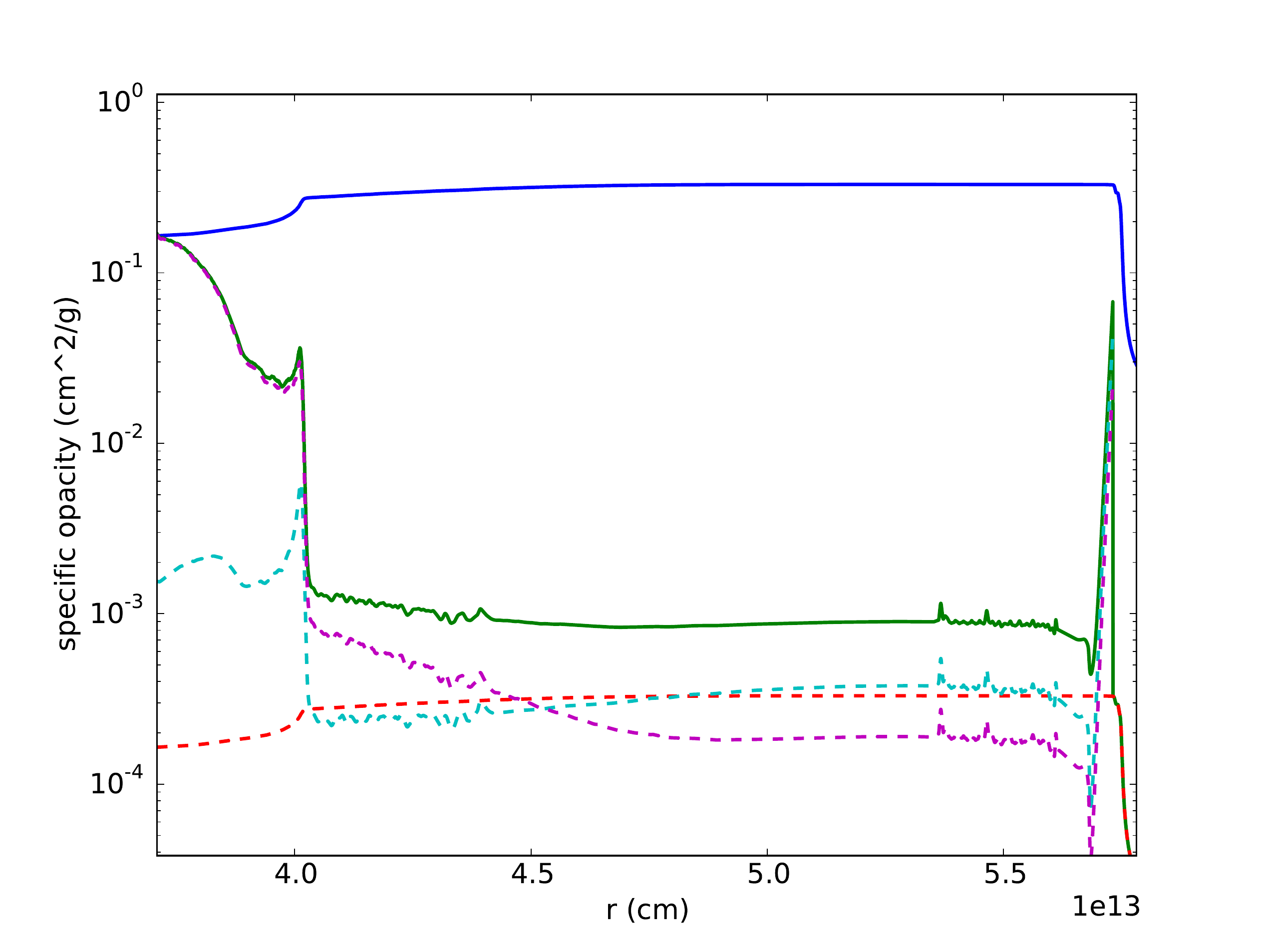}
\caption{ \lFig{f1_opac} Opacity profile for a 15 \Msun RSG near breakout with 5.07\E{50} erg of final kinetic energy (model F15 in \Sect{progenitor}), showing specific opacity $\kappa$. These opacities are calculated using the analytic formulas outlined in \Sect{opac_theory}. Solid lines show total opacity (blue) and total absorptive opacity (green). Dotted lines show absorptive opacity contributions from Compton scattering (red), inverse bremsstrahlung (cyan), and bound-free (magenta). Absorptive opacity is much weaker than scattering opacity in this higher-energy model, and scattering easily dominates total opacity.}
  \end{center}
\end{figure} 

\subsection{Bound-Bound Processes (line opacities)}
\lSect{boundbound} 

Line opacities are not accounted for in these simulations due to the complexity and computational cost. Incorporating them would require either generating tables for the appropriate temperature and densities or manually implementing a much larger opacity network that would be frequency-dependent, negating the work done to place other opacities in simpler grey forms and requiring lengthy multigroup simulations for even the bolometric light curves. 

The effects of line broadening due to velocity shear in the region of the photosphere are also not included in these calculations, but fortunately the shocks considered here are low-velocity as a rule, and at the explosion energies where electron scattering does not dominate envelope velocities are $\sim 500$ km/s. Thus velocity shear effects can be ignored.

To examine this analytic opacity model, tabulated values from the OPAL database were used for comparison. OPAL tables incorporate the effects of lines, as well as many other opacity processes, to a high degree of precision. OPAL can only give total opacity and not separate absorption and scattering opacities, which these simulations require. Still, it is a useful point of comparison. \Fig{opal_comparison} presents opacities as calculated by the analytic formulas discussed here compared to calculations using OPAL for a low-energy explosion near shock breakout. The OPAL opacities are systematically higher; this is expected to be the case as OPAL includes more absorption processes. Thus the opacities in these simulations should be taken as lower bounds. This implies that the peak luminosity and color temperature results from these simulations are upper bounds.

\begin{figure}
\begin{center}
\includegraphics[scale=0.4, clip=true, trim = 25 0 0 0]{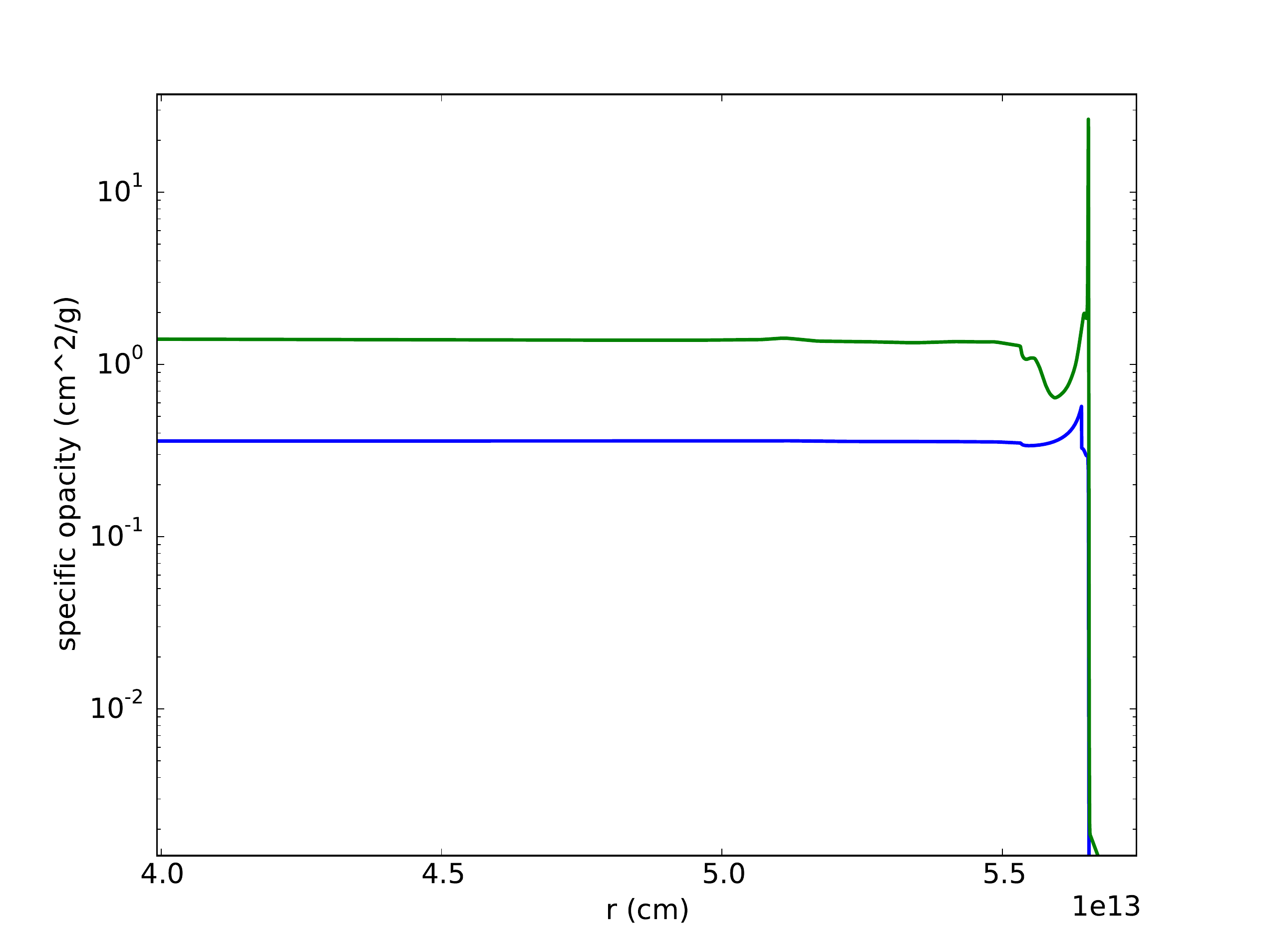}
\caption{\lFig{opal_comparison}Specific opacities as calculated by two different opacity routines, this work (blue) and OPAL (green), for a 15 \Msun RSG near breakout with 1.21\E{49} erg of final kinetic energy (model C15 in \Sect{progenitor}). OPAL opacities are expected to be systematically higher as this work does not account for bound-bound processes and generalizes photoionization to a frequency-independent form. The biggest divergence between the two calculations occurs where the gas temperature drops below 4.5\E{4} K and hydrogen and helium begin to recombine; in this regime the Kramer's Law forms for bremsstrahlung and photoionization used by the analytic model begin to break down.}
\end{center}
\end{figure}

\section{Color Temperature Behavior in VLE SNe}
\lSect{vle_tcol}

Shock breakout theory allows for a difference between \Teff and \Tcol because the spectral shape of the radiation may be set at a chromosphere depth \Rth while the overall energy is set at the photosphere depth \Rp\!, and during breakout these two radii are not necessarily equal. The ratio between \Tcol and \Teff is not set directly by the physics of shock breakout. \citet{nakar10} predicts a ratio 1.8 in red supergiants and 2.1 in blue supergiants. \citet{Rab10}, who consider a low-temperature phase of the supernova akin to conditions during a VLE SN breakout, calculate a ratio in the range 1.1 - 1.8. Many (but not all) numerical simulations of breakout show a ratio of 2-3 between \Tcol and \Teff \citep{eb92, t12, kc78}. It is reasonable, therefore, to ask what kind of ratio can be expected in the VLE case.

In the following analysis it is important to distinguish between the optical depth as measured from the stellar surface, denoted as $\tau'$, as opposed to the optical depth of a single shell, denoted simply $\tau$. In a simulation context a ``shell" is assumed to refer to one cell of a 1D spherical model, each cell having a single value for both opacity and density. $\tau$ is therefore equal to $\kappa\rho\Delta r$. In all following discussions of opacity we also assume both electrons and ions to be at the single gas temperature $T_g$ tracked by CASTRO.

As \citet{nakar10} describe, if a shell of gas can produce sufficient photons to create a blackbody spectrum at its local gas temperature before the breakout radiation diffuses through it, it will set the radiation's color temperature to its gas temperature. The emitted peak color temperature therefore reflects the gas temperature of the outermost shell both a) still coupled to the radiation and b) capable of producing sufficient photons to meet this thermalization criterion. This is not necessarily the radius predicted from optical depth arguments. \citet{nakar10} use this criterion to create the parameter $\eta$, which is the ratio of the number of photons a shell must produce in order to set the radiation to its local gas temperature, divided by the number of photons that can be produced in the appropriate time:
\begin{align}
\eta &= \frac{n_{BB}}{t_{s}\dot{n}_{\gamma}}
\end{align}
where $t_s$ is the time radiation spends in the shell, which is at minimum $\Delta r/c$. While the photons are propagating in the diffusive regime, the time they spend in each shell is $\tau\Delta r/c$. If $\eta < 1$, the shell can produce sufficient photons and it will thermalize the radiation. The emitted peak color temperature therefore reflects the gas temperature of the outermost shell still coupled to the radiation and having $\eta < 1$, which is here called the chromosphere $R_c$. In the limit that radiation spends a full diffusion time in each cell, \citet{nakar10} notes that the condition $\eta = 1$ can be alternately expressed as each photon experiencing on average one absorption in the shell, placing the chromosphere at the location where $\tau_a\tau \approx 1$.

Shock breakout begins when $\tau' \approx c/v_s$ and the radiation can escape the star ahead of the shock. Opacity drops steeply in the {\red pre-shock region} and once radiation can travel ahead of the shock it will escape from the rest of the star without further interaction. Thus the condition $\tau' = c/v_s$ is also the condition $\tau = c/v_s$ and the corresponding shell is the photosphere of breakout.\footnote{Assuming that the medium around the star is transparent to the breakout flash. This is assumed to be true in the models considered here, but in the case of stars with dense CSM/strong stellar winds, it may not be, and the photosphere may be external to the star entirely. The modeling of breakout in stars with complex CSMs is of great interest, but well beyond the scope of this work.} If $\tau \approx \tau_a$, then $\eta = 1$ is $\tau_a^2 \approx 1$ and $R_c$ should now be at $\tau_a \approx 1$. However, the breakout criterion is now $\tau_a \approx c/v_s$. As $c/v_s > 1$, this would place $R_p$ at a higher optical depth, i.e. a smaller radius, than $R_c$; but $R_c$ cannot exceed $R_p$ since past $R_p$ the radiation no longer spends a diffusion time in each cell and on average is not expected to interact. Thus as long as $\tau_a \approx \tau$, $R_c \approx R_p$ and $T_c \approx T_{ef}$. 

A typical shock velocity near the surface in a blue supergiant progenitor with a standard-energy ($>$ 0.6 B) shock model is 1.5\E{4} km/s. At this velocity breakout would occur at the radius where $\tau = c/v_s = 20$. If $\tau_a \ge 0.05$ at the same radius, the spectrum will remain in thermal equilibrium and $R_c = R_p$; otherwise the radiation will drop out of thermal equilibrium before reaching the photosphere and $\Tcol > \Teff$. A typical value for $\tau_a$ at the shock front in the SN87A case is of order $10^{-3}$. Note that since the temperature is changing rapidly just behind the shock, $R_c$ does not have to be much smaller than $R_p$ to give a significant $T_c > T_{ef}$.

The primary differences in the VLE SNe case are the significantly lower gas temperatures and shock velocities. Absorptive opacity is higher due to the low-T effects on opacity as discussed in \Sect{opac_theory}. At the greater optical depth a typical photon experiences many more collisions (essentially $\tau^2$) and will thus have more opportunity to thermalize even in the presence of a small absorptive opacity. At $v = 1\E{3}$ km/s breakout would occur at $\tau \approx 300$ and $R_c = R_p$ if $\tau_a \ge 0.003$ at that location. This is an easier condition to meet, especially in a regime with significant contributions from inverse bremsstrahlung, and values of $\tau_a$ in this region are of order $10^{-2} - 10^{-1}$. 

To further illustrate the difference with energy, consider a simplified model implementing only electron scattering for both \chiR and \kP, with the absorptive component represented by a prescribed ratio $\epsilon = \kP/\chiR$ (see \Sect{freefree}). Thus $\tau_a = \epsilon\tau$ and the criterion for color temperature becomes $\epsilon\tau^2 = 1$, so the color temperature is set at the depth $\tau = 1/\sqrt{\epsilon}$. Then, to achieve \Tcol = \Teff, the standard-energy example requires $\epsilon > 2.5\E{-3}$ while the VLE example would only require $\epsilon > 1\E{-5}$.

The color temperature can therefore be expected to converge with the effective temperature as the shock temperature goes down. This results in a breakout that is brighter in the
IR and optical windows than would be expected because the bolometric correction is smaller. The observational impacts of this effect are discussed in \Sect{vle_ir}.

\subsection{Velocity Effects}
\lSect{velocity_term}
{\red Initial attempts to simulate shock breakout in SN 1987A produced unexpectedly high radiation color temperatures, particularly at the shock front; gas temperature remained steady, but \Tcol increased by an order of magnitude. This discrepancy was ultimately resolved by implementing higher, more accurate absorption opacities. It is worth considering why the simulation results were so sensitive to changes in absorption opacity, as an illustration of the necessity of considering opacity details. Examining the different terms affecting the spectrum in the flux-limited diffusion approximation provides the solution.} During breakout, two terms in the radiation-hydro equations dominate the spectrum: an advection term in frequency space and the matter-radiation coupling term. The multi-group treatment of flux-limited diffusion takes the form:

 \begin{align}
\pd{E_g}{t} + \div{\left(\frac{3 - f_g}{2}E_g\vel\right)} - \vel \cdot \grad\left(\frac{1-f_g}{2}{E_g}\right) &= \\
c\kappa_g(\alpha T^4 - E_g)  + \div{\left(\frac{c\lambda}{\chiR}\grad{E_g}\right)} &\\
+ \int_g\pd{}{\nu}\left[\left(\frac{1-f}{2}\div{\vel} + \frac{3f-1}{2}\hat{n}\hat{n} : \grad\vel\right)\nu E_\nu\right] &\\
- \frac{3f -1}{2}E_g\hat{n}\hat{n} : \grad\vel
\end{align}
Consider the optically thick limit where the gas has the most influence on the radiation, in which case $f_g = 1/3$ and all $\hat{n}\hat{n}$ terms (representing free streaming radiation) disappear.
\begin{align}
\pd{E_g}{t} + \frac{4}{3}\div{E_g\vel} - \frac{1}{3}\vel \cdot
\grad\left({E_g}\right) &= \\
c\kappa_g(\alpha T^4 - E_g)  + \div{\left(\frac{c}{3\chiR}\grad{E_g}\right)} &\\
+ \frac{1}{3}\int_g\pd{}{\nu}\left[\left(\div{\vel}\right)\nu E_\nu\right] 
\end{align}
On the left-hand side, the first term in the equation represents the quantity to be solved for, the change in group energy with time. The second and third terms represent energy transfer by bulk fluid motion. The three terms on the right-hand side respectively represent energy exchange via absorption/emission, diffusion of radiation, and energy transfer between frequency groups.

{\red This equation can be interpreted as an advection equation in frequency space (see \citet{castro3} for details). The ``sound speed" in this medium includes a term dependent on the divergence of velocity, $\nabla \cdot \vec{\vel}$. In practice, this means that as the gas is compressed by the shock, a velocity divergence arises that shifts the spectrum towards the blue. This term can cause the radiation spectral temperature to diverge from the gas temperature even in optically thick regions. In particular, if this term is the only one considered, the spectrum will harden significantly just before breakout as the shock reaches the stellar atmosphere and increases in velocity. The magnitude of this effect is directly linked to the velocity at the shock front - higher velocity leads to greater divergence and more hardening of the spectrum. 

Conversely, the absorption-emission source exchange term, which depends on absorptive opacity, will tend to equilibrate the radiation and the matter, so in the case of shock breakout it will cool the radiation. The relative strength of these two terms determines the ultimate color temperature. If the source exchange term dominates, \Tcol will follow the gas temperature. If the advection term dominates, however, \Tcol can be significantly higher. Therefore the relative timescales of these terms must be considered in order to ensure the resulting color temperature has the correct qualitative behavior.}

Let the scale length $L = c/\chiR \vel$ where \chiR is the scattering
opacity. The terms that will alter the spectrum are the frequency
group coupling term $(1/3)(\div{\vel})\nu E_g$, which depends on
velocity divergence, and the absorption/emission energy exchange term
$c\kP(\alpha T^4 - E_r)$. The time scale of the exchange term is the
inverse of $c\kP$\!. The time scale of the velocity divergence term is
the inverse of $\vel / 3L$.
\begin{align*}
\tau_{x}^{-1} &= c\kP\\
\tau_{v}^{-1} = \frac{\vel}{3L} &= \frac{1}{3}\vel\left(\frac{\chiR\vel}{c}\right)\\
\frac{\tau_{v}}{\tau_{x}} = \theta = \left(c\kP\right)\left(\frac{3c}{\chiR\vel^2}\right) &= 3\left(\frac{c}{v}\right)^2\left(\frac{\kP}{\chiR}\right)
\end{align*}
If $\theta > 1$, $\tau_v > \tau_x$, the exchange term dominates, and
the radiation spectral temperature will follow the gas. If $\theta <
1$, $\tau_v < \tau_x$, the velocity term dominates, and the radiation
spectral temperature will increase above the gas temperature in
regions of high velocity divergence, e.g. shock fronts.

For the SN1987A model, $v \sim 15000$ km s$^{-1}$ at breakout.  If $\kP/\chiR$ = $10^{-6}$, $\theta = 0.0012$; if $\kP/\chiR$ increases to $10^{-3}$, then $\theta = 1.2$, implying a significant shift in the spectral behavior of the simulation. \Fig{87a_epsilon} shows the temperatures of three SN1987A simulations run up to the moment of breakout in order to test this hypothesis. Since at these temperatures electron scattering dominates both \kP and \chiR, the opacity network was simplified to only calculate this process, which allows easy variation {\red by changing the parameter $\epsilon = \kP/\chiR$.} Solid lines represent gas temperature and dashed lines represent color temperature. Where the two lines overlap, the radiation and gas are in equilibrium. \footnote{{\red The stair-step appearance of the \Tcol curves is an artifact of the multigroup approximation, where radiation is approximated as a set of groups each corresponding to a range of frequencies. A color temperature calculation done by selecting the group with the highest energy density and applying Wien's Law to its central frequency will therefore show discrete changes in value corresponding to the boundaries of the frequency groups.}} These simulations used 64 frequency groups spaced logarithmically in the range 1\E{15} - 1\E{18} Hz.

{\red The red and green dashed lines, representing color temperature in simulations with $\epsilon = 10^{-6}$ and $10^{-4}$ respectively, both spike above the gas temperature at the shock front
- the region of highest velocity divergence - and then diffuse forward. By contrast in the $\epsilon = 10^{-3}$ simulation (blue) the color temperature follows the gas temperature through the hydrodynamic shock. Thus, the anomalously high color temperatures in the initial simulations, which were run with $\epsilon = 10^{-6}$, are due to velocity effects. Changing this ratio to a more realistic $\epsilon \sim 10^{-3}$ brought \Tcol in line with previous simulations, as will be discussed in \Sect{87a}.}

Shock front velocities in the red supergiant models, especially the VLE explosions modeled here, are much slower than in SN 1987A. Velocities at breakout range from 80 to 1200 km s$^{-1}$ for RSG15. and 150 to 500 km s$^{-1}$ for RSG25. Even for $\kP/\chiR$ = $10^{-6}$ the exchange term dominates in most RSG cases. Velocity effects are therefore unlikely to play a role in VLE SNe breakouts.

\begin{figure}
\includegraphics[scale=0.4, clip=true, trim = 25 0 0 0]{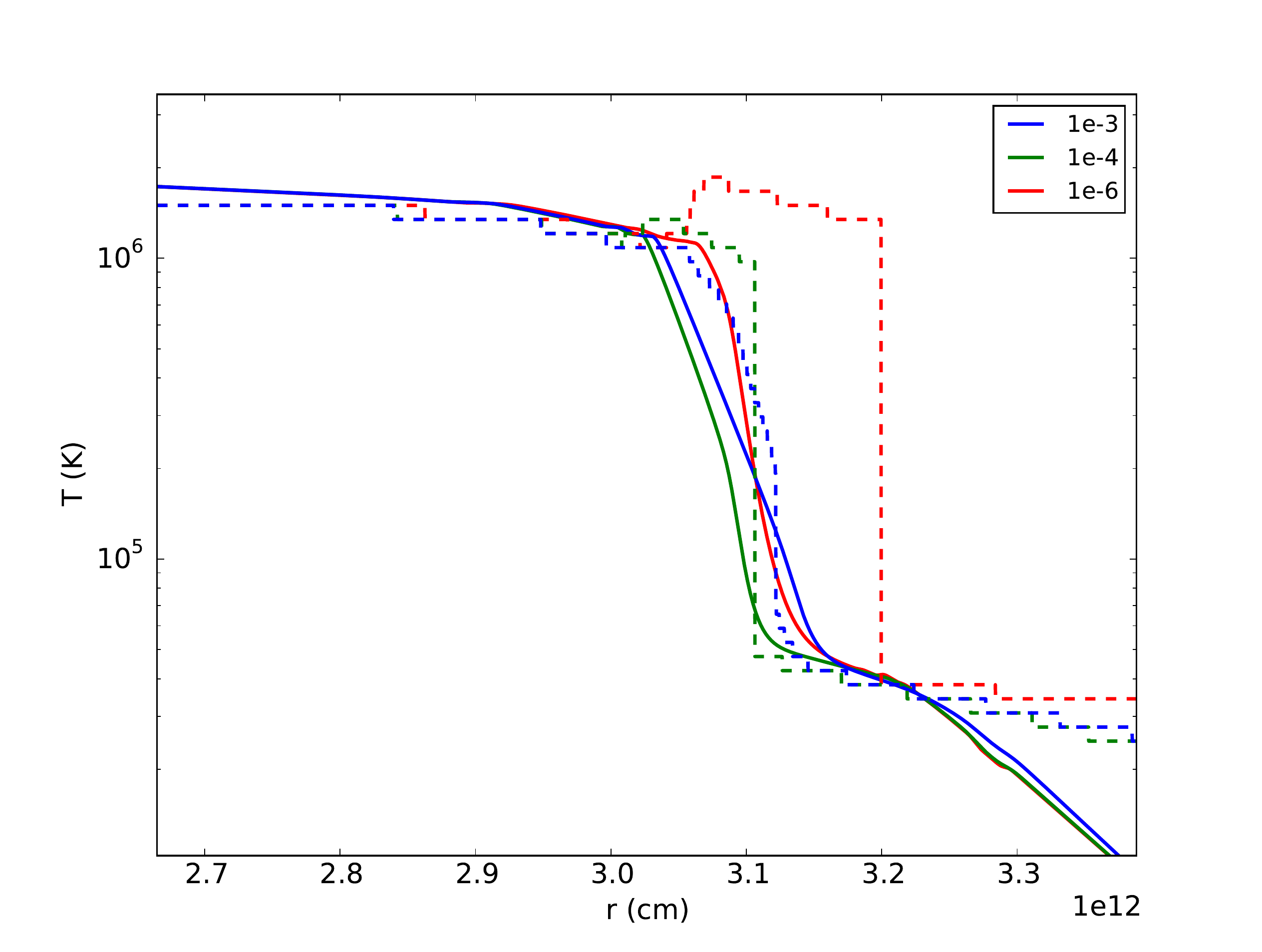}
\caption{ \lFig{87a_epsilon} Temperatures in SN1987A shock breakout
  for the 1.0 B model with three choices of $\epsilon = 1\E{-6},
  1\E{-4}, 1\E{-3}$ (red, green, blue). Solid lines show gas
  temperature and dashed lines show radiation spectral temperature as
  estimated by measuring which frequency group has the highest energy
  density. At low $\epsilon$ (red, green), the high velocity
  divergence at the shock front causes the spectral temperature to
  increase above the gas temperature even when the material is still
  optically thick.  As $\epsilon$ increases, radiation-gas energy
  exchange counteracts the effects of the velocity divergence and the
  spectral temperature stays in equilibrium with the gas temperature
  longer. The stair-step nature of the color temperature curves is an
  artifact of the multigroup approximation, where radiation is
  approximated as a set of groups each corresponding to a range of
  frequencies. These calculations are shown very close to the moment
  of breakout, when radiation has just begun to diffuse out from
  behind the shock.}
\end{figure} 

In cases with high velocity, low metallicity, or both, this effect may still come into play. Supernovae of more standard energies exploding in very metal poor blue supergiants, as theorized for some models of first-generation stars, might also have unusually high color temperatures. While a full simulated exploration of this space is beyond the scope of this paper, we can qualitatively predict some behavior. Stars with high envelope metallicities and/or low shock velocities will likely show the same 2 - 3 ratio of \Tcol to \Teff as SN1987A, but stars with very low envelope metallicities, such as Pop III stars, might exhibit spectral temperatures dramatically higher than predicted. \citet{danppsn} estimated the shock velocity at breakout in pair SNe in a red supergiant model at $1.5\E{4}$ km/s, similar to the SN1987A model discussed in this section; in blue supergiants this number can be substantially higher. At the temperatures of a standard-energy breakout inverse bremsstrahlung is highly suppressed and metals contribute most of the photoionization opacity; in a low-metal star total absorption opacity may therefore be quite low. At high velocities and low absorptive opacity the velocity divergence discussed here can easily dominate the exchange term and make the spectrum much hotter than opacity arguments would predict.

\section{Verification and Validation of CASTRO: SN1987A}
\lSect{87a}

\begin{deluxetable*}{cccccccccc}
\tablewidth{0pt}
\tablecaption{SN1987A Breakout Model Results}
\tablehead{
\colhead{Model} &
\colhead{Star} &
\colhead{$v$\tablenotemark{a} (km/s)} &
\colhead{KE\tablenotemark{b} (ergs)} &
\colhead{$L_{peak}$ (erg/s)} &
\colhead{$\Delta t$\tablenotemark{c} (s)} &
\colhead{Max \Teff (K)} &
\colhead{Max \Tcol (K)} &
\colhead{$\nu_{peak}$ (Hz/\AA)}}
\startdata
1.0 B & SN1987A & 2.0\E{4} & 1.08\E{51} & 6.77\E{44} & 70.7 & 5.13\E{5} & 1.1\E{6} & 6.47e16\\
2.3 B & SN1987A & 3.2\E{4} & 2.01\E{51} & 1.53\E{45} & 45.0 & 6.29\E{5} & \nodata & \nodata
\enddata
\tablenotetext{a}{Velocity at breakout.}
\tablenotetext{b}{Kinetic energy at breakout.}
\tablenotetext{c}{Full-width half-max of travel-time-corrected light curve.}
\lTab{87aresults}
\end{deluxetable*}

Because CASTRO's MGFLD module is new, it was important to first verify its capabilities for this sort of simulation. A frequently modeled shock breakout event is that of SN1987A, which has been previously simulated by \citet{eb92} (EB92) with the VISPHOT code and \citet{t12} (T12) in 2012 with the STELLA code. SN1987A also has upper limits on the temperature and luminosity of its shock breakout that were derived after the fact from observations of the ionization of the surrounding gas \citep{fransson89, lundqvist96}. The progenitor chosen was an 18 \Msun blue supergiant with a radius of $3.2\E{12}$ cm, shown in \Fig{87a_rhoT}, the same model studied in EB92. This model was also studied in \citet{Suk15}. Two different explosion energies were sampled. One model at 1.0 B was close to the observed value, near $1.3 \times 10^{51}$ erg \citep{Arn89} and the other model, at 2.3 B, was chosen to match the second simulation in EB92. The high temperatures in SN1987A's breakout ensure that electron scattering opacity is dominant.

\begin{figure}
\begin{center}
\includegraphics[scale=0.4, clip=true, trim = 25 10 70 50]{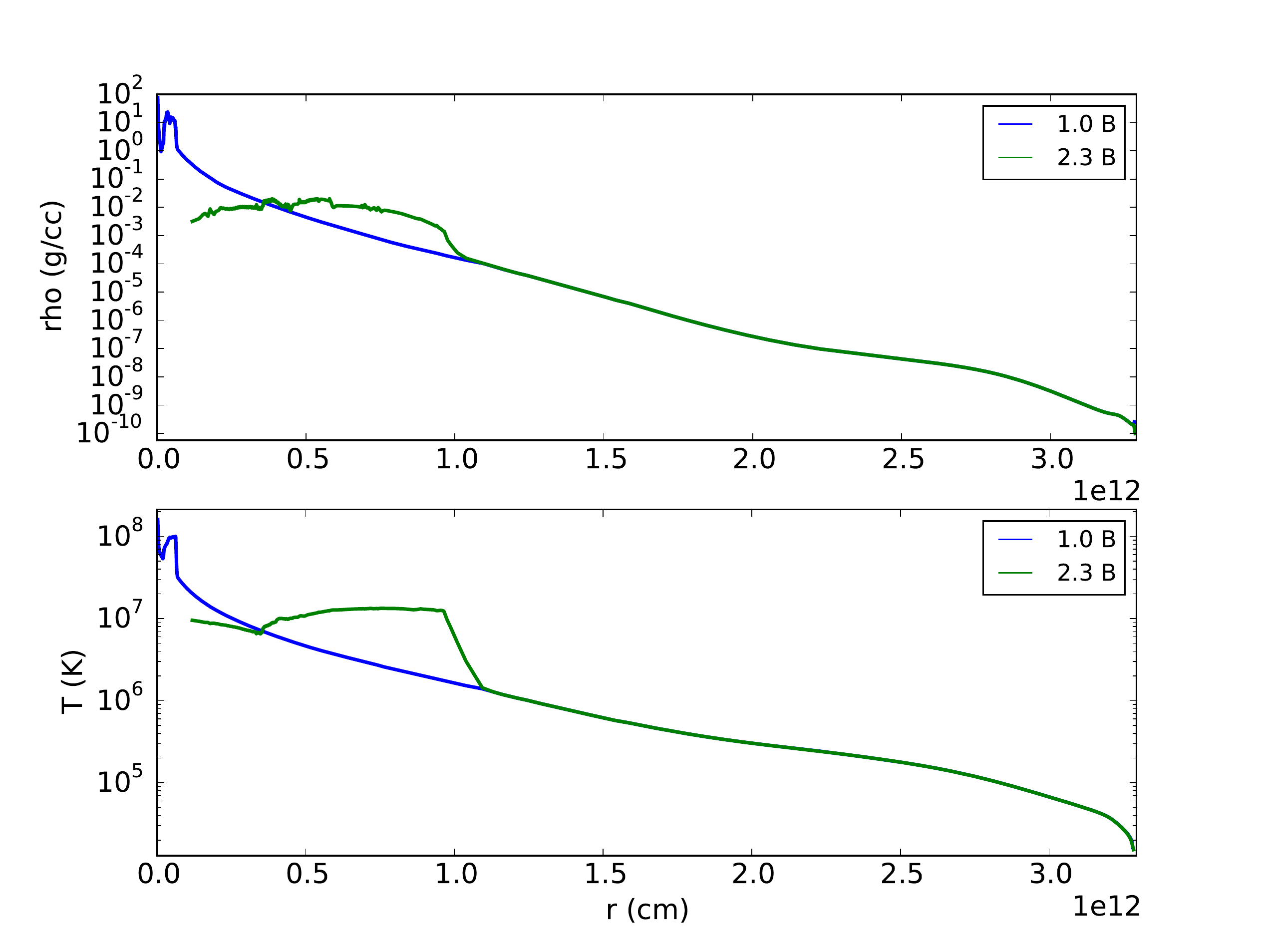}
\caption{ \lFig{87a_rhoT} Density and temperature profiles for the
  SN1987A progenitor at the time the calculation was linked from the
  KEPLER code to CASTRO for two different explosion energies, 1.0 B
  (blue) and 2.3 B (green).}
  \end{center}
\end{figure} 

The 1.0 B explosion was simulated in both single-group (grey radiation) and multigroup mode; the 2.3 B explosion was studied only in single-group. The results are shown in \Fig{87a_lc} which gives the bolometric light curve, and \Fig{87a_spec} which gives the
spectrum of the 1.0 B model at peak $T_c$. Peak \Tcol will occur near
the very beginning of the breakout, while the bolometric luminosity
will rise on a longer timescale; in SN1987A the peak luminosity occurs
about 100 seconds after peak color temperature. Bremsstrahlung and
photoionization opacities are negligible in this energy regime and are
not included. Comptonization is included and the ratio \kP/\chiR has
been set to $5\E{-3}$ in order to ensure the source exchange term is
dominant (\Sect{velocity_term}). The bolometric light curve is similar
to past studies with a luminosity that peaks around $10^{45}$ erg/s,
indicating a high-energy explosion. The breakout light curve has a
width of $\sim 100$ s, indicating a small stellar radius. Peak
effective temperature occurs at peak luminosity, and the ratio of peak
temperatures \Tcol/\Teff $= 1.1\E{6} $ K$/5.1\E{5} $ K = 2.2, matching
previous calculations. The 2.3 B explosion has a higher and sharper
peak than the 1.0 B. A summary of the results is given in
\Tab{87aresults}. CASTRO's results therefore match both prior
simulations and theoretical predictions for the test case of SN1987A.

\begin{figure}
\begin{center}
\includegraphics[scale=0.4, clip=true, trim = 25 10 70 50]{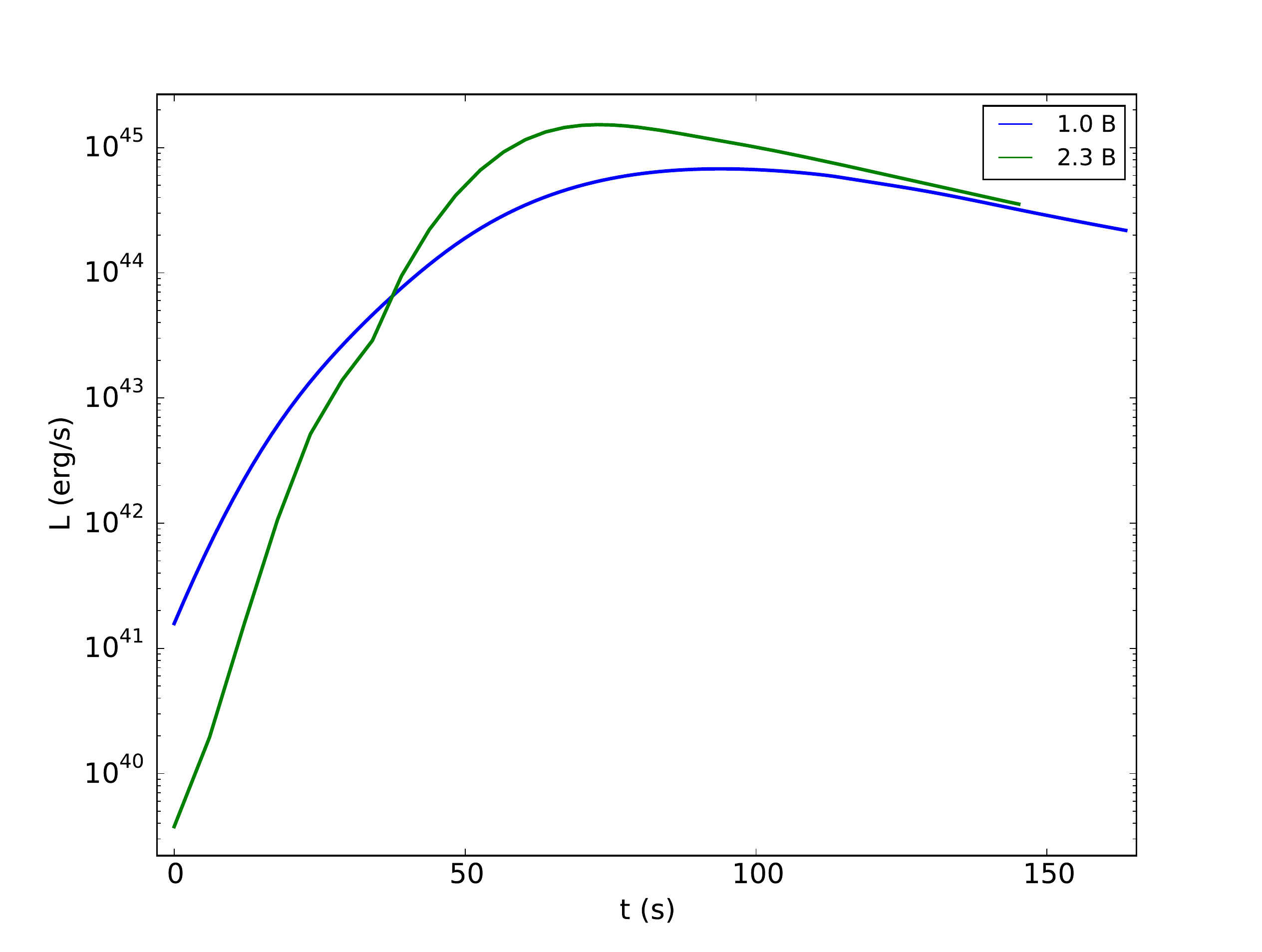}
\caption{ \lFig{87a_lc} Bolometric light curves for shock breakout in
  SN1987A calculated for two different explosion energies using
  CASTRO, 1.0 B (blue) and 2.3 B (green). The higher-energy breakout
  is significantly brighter and shorter. The curves have been
  arbitrarily shifted in time to overlay at peak for ease of
  comparison.}
\end{center}
\end{figure} 

\begin{figure}
\begin{center}
\includegraphics[scale=0.4, clip=true, trim = 25 10 60 40]{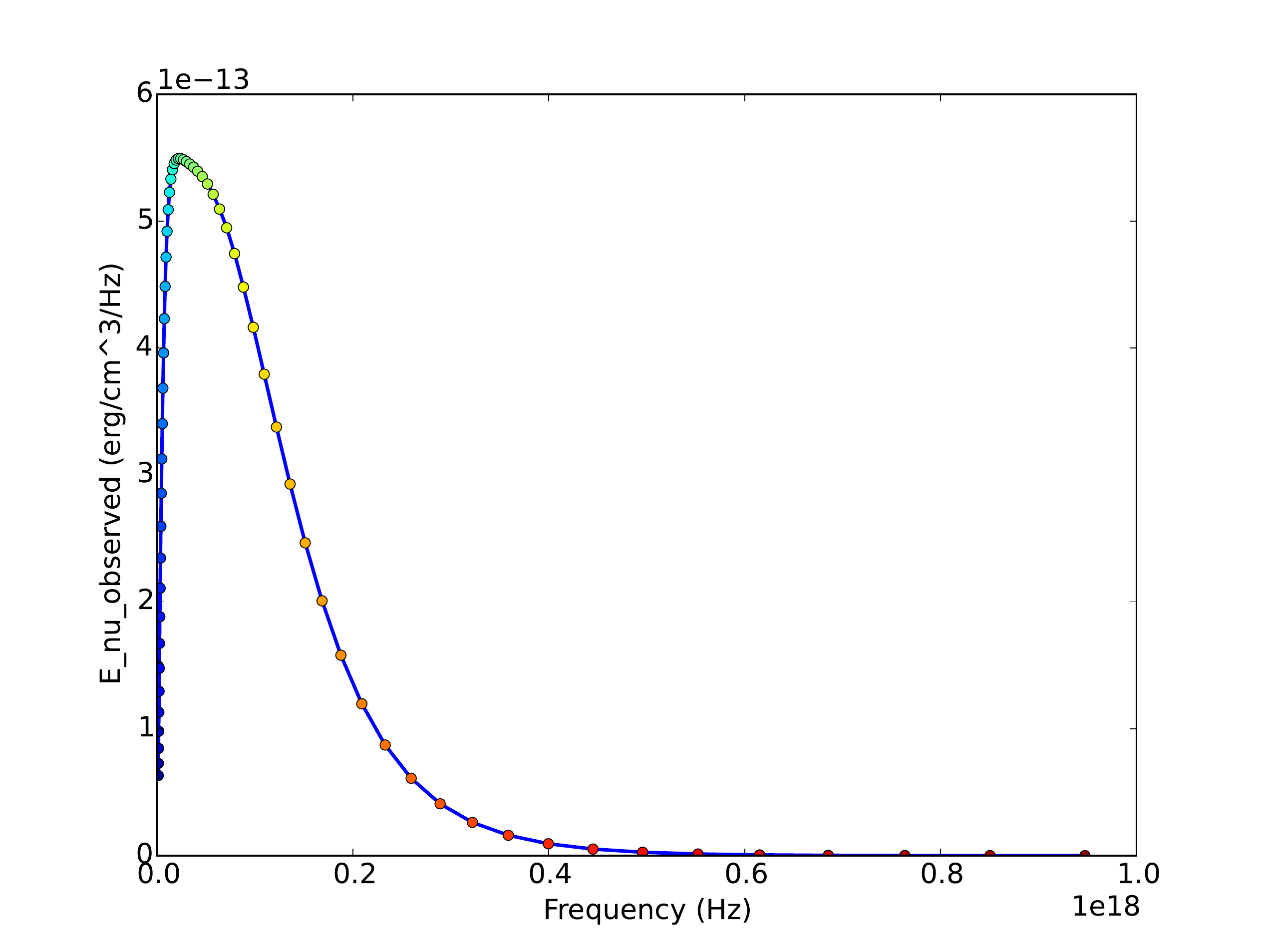}
\caption{ \lFig{87a_spec} Spectrum for shock breakout in SN1987A for a
  1.0 B explosion calculated using CASTRO and sampled at peak color
  temperature. Circles mark the centers of frequency
  groups. Sixty-four groups were used in this calculation. This
  spectrum has a blackbody form and an effective temperature of \Teff
  = 5.41\E{5} K, but applying Wien's Law to the peak frequency gives a
  color temperature \Tcol = 1.1\E{6} K.}
\end{center}
\end{figure}

\section{Simulation Setup For the Main Study}
\lSect{setup}

\subsection{Progenitors}
\lSect{progenitor}

Based on a number of recent surveys \citep[e.g.,][]{Oco11,Pej15,Suk15}
the presupernova stars that most commonly produce black holes in stars
that still retain their envelopes had main sequence masses in the
range 15 to 35 \Msun\!. Especially prolific in black hole production
may be the stars 18 to 26 \Msun \citep{horiuchi11, ugliano12,
  justinnse, Tcompactness, kochanekBH, clausenBH}. In order to sample
this range, two progenitor stars are used here, both red supergiants,
with main sequence masses 15 \Msun (RSG15) and 25 \Msun (RSG25). Their
presupernova structures are shown in \Fig{rsg15_rhoT},
\Fig{rsg25_rhoT}, and \Tab{models}. Since the radiation produced by
breakout depends chiefly on the shock wave energy, which will be
varied, and the presupernova radius, which changes only by about a
factor of two in the mass range of interest, these two models should
suffice.  The models here are taken from \citet{Woo07} and are very
similar to those more recently explored by \citet{Tcompactness} and
\citet{Suk15}.

These stars were evolved using the KEPLER code \citep{Wea78,Woo02} from ignition on the main sequence to the time of collapse. An artificial shock of variable energy was initiated by first extracting the iron core and then dumping energy in the bottom 10 zones. Since the calculations here focus on the early light curve and its plateau and do not depend upon nucleosynthesis or details of how the explosion was powered, this simple approach should suffice.

It is important to note that the energy deposited, the shock energy at
the time of breakout, and the final kinetic energy at infinity of the
ejecta are three different quantities. Not all energy deposited
becomes kinetic, and the shock energy at breakout is not the same as
the total energy, which includes both positive internal and negative
gravitational energy. Both the kinetic energy at breakout, which is
relevant for breakout, and the final kinetic energy, which is relevant
for the later light curve, are given in \Tab{rsgresults}.

In most core-collapse simulations the properties of the pre-collapse
stellar atmosphere are irrelevant, since by the time the explosion is
seen, the layers that governed breakout have already become
transparent.
Care must be taken to treat it accurately in shock breakout
studies. Additional modifications were therefore made to the
progenitor models' atmospheres. These modifications are discussed in
\Sect{rsg_atm}.

Three calculations were done for each case: a precise breakout
simulation using CASTRO (\Sect{breakoutrsg}); a simplified breakout
simulation using KEPLER for comparison purposes; and a precise
calculation of the later plateau-stage light curve using KEPLER
(\Sect{kepler_comp}). KEPLER uses only single temperature flux-limited
radiation transport, but as long as a KEPLER progenitor model is
mapped into CASTRO at a time when the material is still very optically
thick and the radiation is fully thermalized, no loss of precision
occurs. Breakout results from KEPLER give information on the
bolometric luminosity and effective emission temperature that is
useful to compare with CASTRO, and with adequate zoning, give
reasonable estimates. These results are discussed further in
\Sect{kepler_comp}. Only the CASTRO multigroup calculation gives
information on the color temperature and spectrum.

Because the energy in the shock at breakout depends only on the local
velocity structure in the hydrogen envelope, models for the 25
\Msun\ case were generated by simply multiplying the velocities in
KEPLER model {\red C25} by a constant factor before inputting them
into CASTRO. All RSG15 models were calculated individually using
KEPLER.

\begin{deluxetable*}{cccc}
\tablecaption{Presupernova Star Parameters}
\tablehead{
\colhead{Star} &
\colhead{Final Mass ($\Msun$\!)} &
\colhead{He Core Mass ($\Msun$\!)} &
\colhead{Radius (cm)}}
\startdata
SN1987A & 15 & 8 & 3.2\E{13}\\
RSG15 & 12.79 & 4.27 & 6\E{13}\\
RSG25 & 15.84 & 8.20 & 1.07\E{14}
\enddata
\lTab{models}
\end{deluxetable*}

\begin{figure}
\includegraphics[scale=0.4, clip=true, trim = 25 0 0 0]{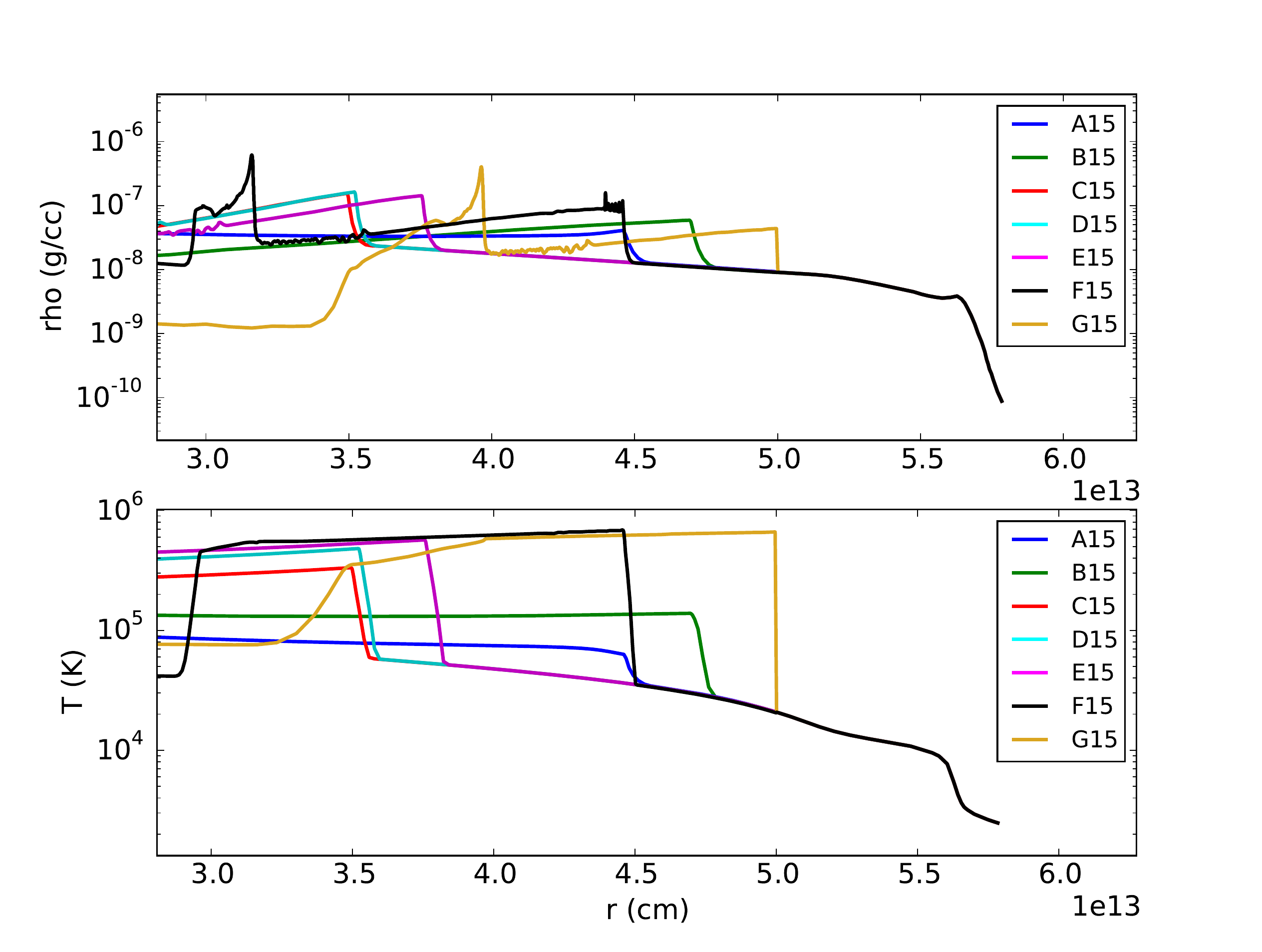}
\caption{ \lFig{rsg15_rhoT} Density and temperature profiles for RSG15
  at the time the calculation was linked from the KEPLER code to
  CASTRO for seven different explosion energies. The shock energies
  increase alphabetically. The time, and hence radius of the link was
  arbitrary, but sufficiently early and deep in that the shock was
  still in very optically thick regions of the star. The KEPLER zoning
  was relatively coarse with only 40 zones external to $5 \times
  10^{13}$ cm. The surface structure is (or should be) unchanged since
  the shock wave was launched and is identical to the pre-supernova
  stellar atmosphere. Because of this coarse zoning and crude surface
  physics, the effect of using a model atmosphere was explored in
  \Sect{rsg_atm}.}
\end{figure} 

\begin{figure}
\includegraphics[scale=0.4, clip=true, trim = 25 0 0 0]{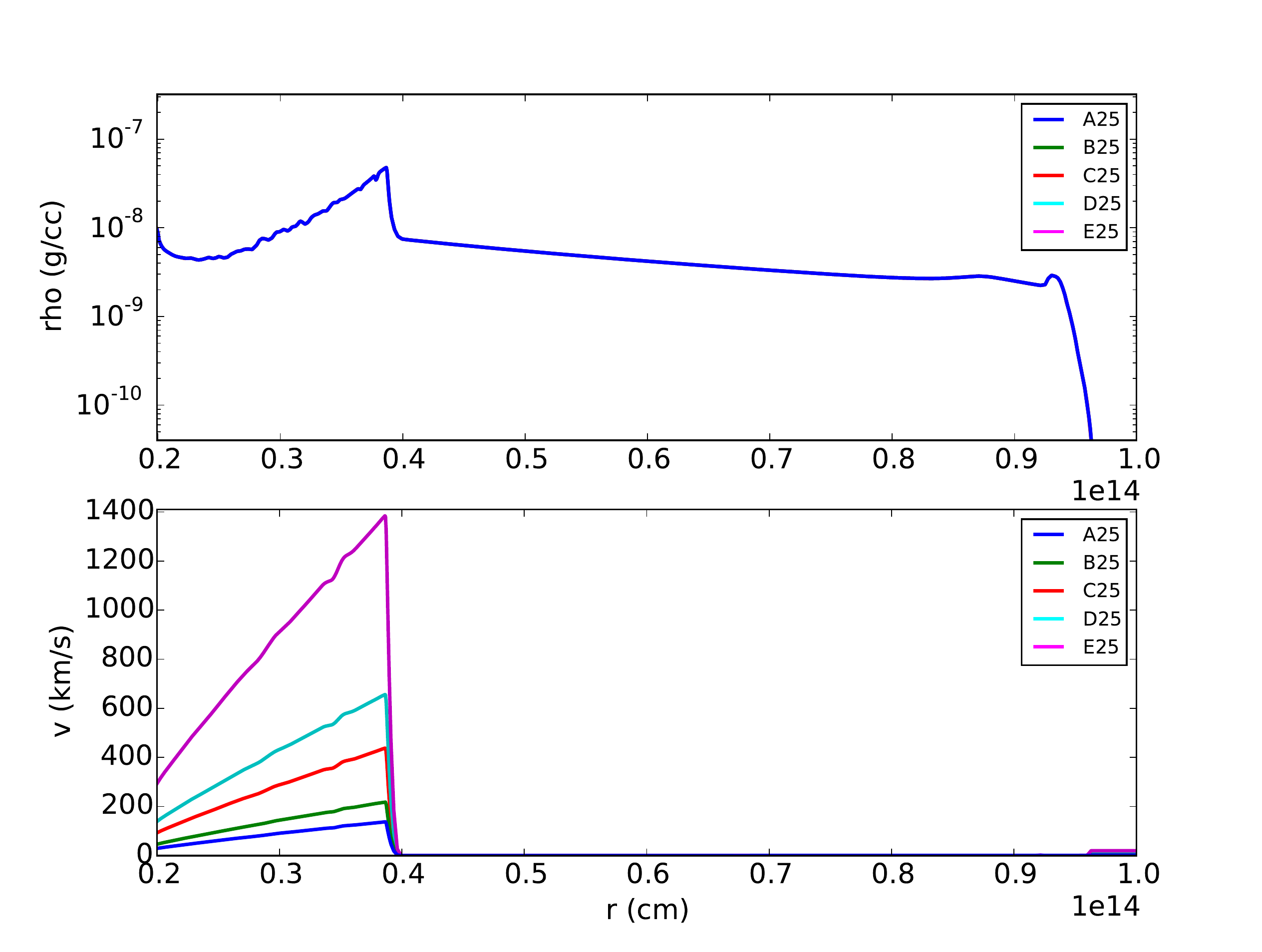}
\caption{ \lFig{rsg25_rhoT} Density and velocity profiles for RSG25 at
  link time from KEPLER to CASTRO. RSG25 models with varying energies
  were produced by multiplying the velocities in a single RSG25 model
  (here designated C25) by a constant factor at link time. Density and
  temperature profiles were assumed to be the same.}
\end{figure}

\subsubsection{Choice of Energies}

Explosion energies in the range $10^{47} - 10^{50}$ ergs were explored for red supergiants. The lower bound on this range is set by the approximate binding energy of the hydrogen envelope. Even the recombination of the hydrogen and helium in the envelope gives this much energy and it is difficult to imagine the ejection of most of the envelope with less energy. This also gives very low velocity and a long transient that might be difficult to detect or be easily confused with other sources. This sort of event might result from nuclear instabilities in the core during oxygen and silicon burning. For example, the stronger silicon flashes studied in stars of near 10 \Msun by \citet{Woo15} imparted a kinetic energy to the envelope of $\sim5 \times 10^{49}$ erg, but the weaker flashes imparted far less, down to a few $\times 10^{47}$ erg. Unfortunately those very weak shocks did not make it to the surface before the core collapsed and a new shock was launched. A shock of $3 \times 10^{48}$ erg did reach the surface though. \citet{waves2} estimate that convection in presupernova stars can drive shocks with $10^{46} - 10^{48}$ ergs of energy. Energies of 10$^{47}$ - 10$^{48}$ erg were also seen in the failed supernovae studied by \citet{nad80} and \citet{unnova}.

``Low mass'' pulsational pair instability supernovae (main sequence masses near 70 - 80 \Msun) can also give very weak shock waves due to the late onset and weak nature of the pulsations (Woosley 2016, in prep.) During the last few hours of the star's life, pulsations in the oxygen burning shell drive strong sounds waves through the helium core that steepen into shocks at the base of the hydrogen envelope. By the time the shock reaches the stellar surface, at 1 -  2 $\times 10^{14}$ cm, the shock energy has degraded and ejection speeds are only a few hundred km s$^{-1}$ or less, corresponding to kinetic energies $\sim10^{47} - 10^{49}$ erg. Slightly more massive stars give pulses from 10$^{49} - 10^{50}$ erg.

The upper bound to the energy range studied is set by the lowest energy supernovae that have already been detected. The Crab supernova for example, is thought to have had an energy of near 10$^{50}$ erg \citep{Yan15}. Although light curves and breakout above $\sim 1\E{50}$ erg have already been studied \citep{tominagatypeII, t12, danppsn, whalenppsn, eb92}, models F15 (0.5 B) and G15 (1.2 B) are included here to study the transition between standard-energy and low-energy behavior.

\subsubsection{Stellar Atmospheres} 
\lSect{rsg_atm}

The KEPLER code is designed to study the internal structure and nucleosynthesis of stars and supernovae and does not treat the outer stellar atmosphere very carefully. However shock breakout is fundamentally an atmospheric phenomenon, since its properties are governed by the critical thick-to-thin transition region near the photosphere. In the SN1987A models, this atmosphere is on the order of $1\E{12}$ cm thick; in the RSG models it is $\sim 6\E{12}$. The effect of variations in the stellar atmosphere was tested by running two different simulations of shock breakout in SN 1987A. The first model had its atmosphere replaced by a power law fit of the form {\red log$_{10} \, \rho = \alpha_1\mathrm{log}_{10}\, r + \beta_1,\ \mathrm{log}_{10}\, T = \alpha_2\mathrm{log}_{10}\, r + \beta_2$, }where $\alpha, \beta$ were determined by fitting the original KEPLER data. The fits gave $\alpha_1 = -42.1,\ \beta_1 = 518.0,\ \alpha_2 = -26.4,\ \beta_2 = 334.5$. The second model then had its atmosphere replaced by the same power law form using $\alpha_1/2, \alpha_2/2$, generating a significantly less steep gradient. The resulting atmospheric gradients are shown in \Fig{87a_atm}. As can be seen in \Fig{87a_atm_lc}, the different atmospheres result in quite different breakout profiles. The two breakouts have the same integrated energy, since that energy is being released from the shock wave itself, but the more slowly varying atmosphere influences the time scale and temperature of the observed transient. A shallower atmospheric gradient creates a wider transition region and thus a breakout that is longer lasting and correspondingly less luminous at peak, while a steeper gradient creates a harder, faster transient.

The true atmosphere of SN 1987A's progenitor is not expected to vary this much between presupernova models, but the comparison does demonstrate that differences in the atmospheric gradient can produce corresponding and significant differences in the resulting light curves. During KEPLER's modeling of the propagation of the shock waves from the core to the CASTRO link point, physics changes designed to improve simulation of the shock wave structure can lead to secondary responses in the atmosphere. Since explosions at different energies take different amounts of time to reach the surface, this can also lead to variations between different models in the same pre-supernova progenitor. These divergences are clearly unphysical and modifications to the CASTRO input models are therefore required.

\begin{figure}
\includegraphics[scale=0.4, clip = true, trim = 25 0 0 0]{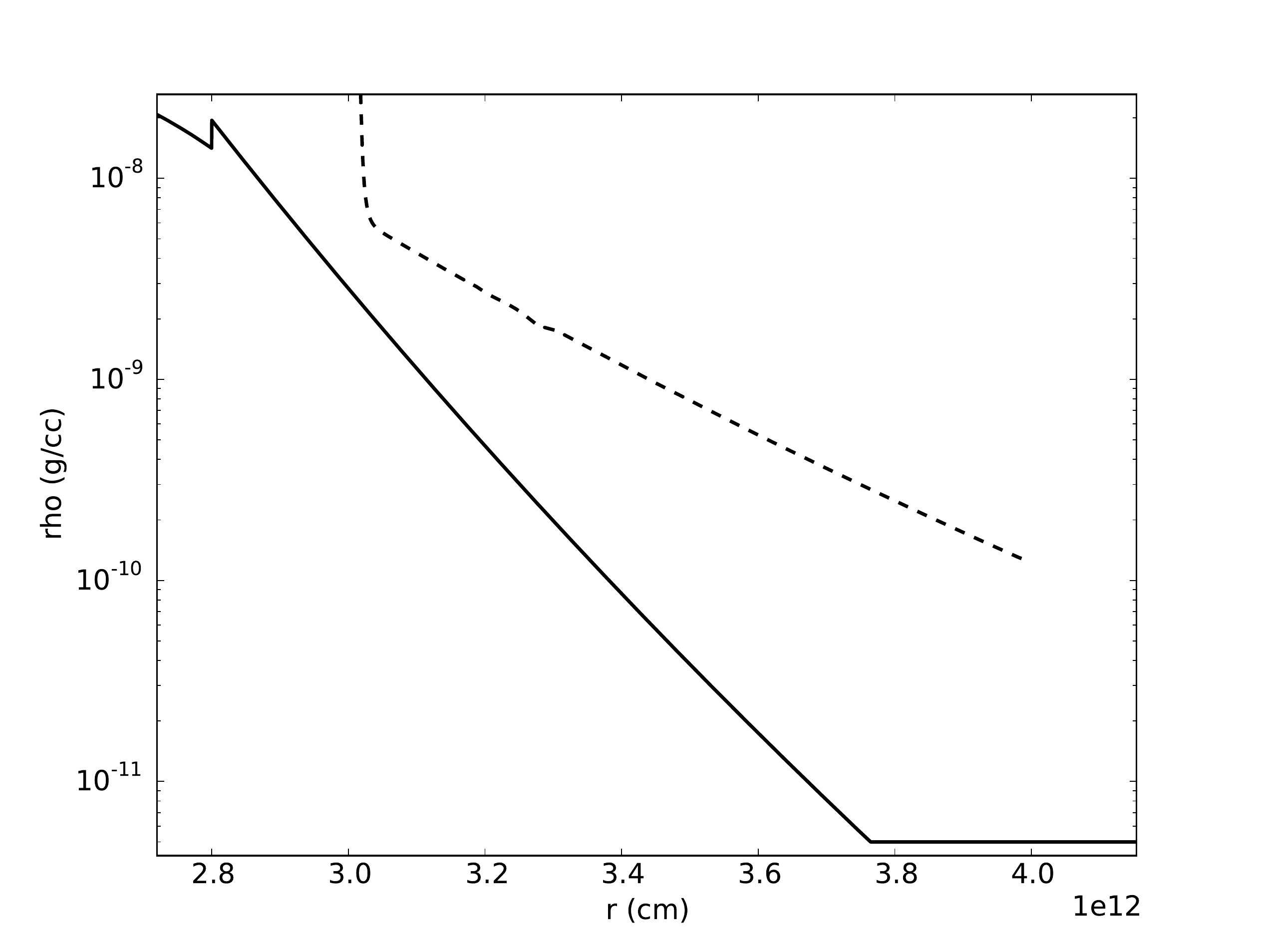}
\caption{ \lFig{87a_atm} Density of SN1987A presupernova model with
  two different stellar atmospheres applied: a power law fit to the
  initial KEPLER model of the form $\rho = \alpha_1\,
  \mathrm{log}_{10} \, r + \beta_1$ (solid), and the same power law
  fit made shallower by using $ \rho = (\alpha_1/2)\,
  \mathrm{log}_{10} \, r + \beta_1$ (dashed).}
\end{figure}

\begin{figure}
\includegraphics[scale=0.4, clip = true, trim = 25 0 0 0]{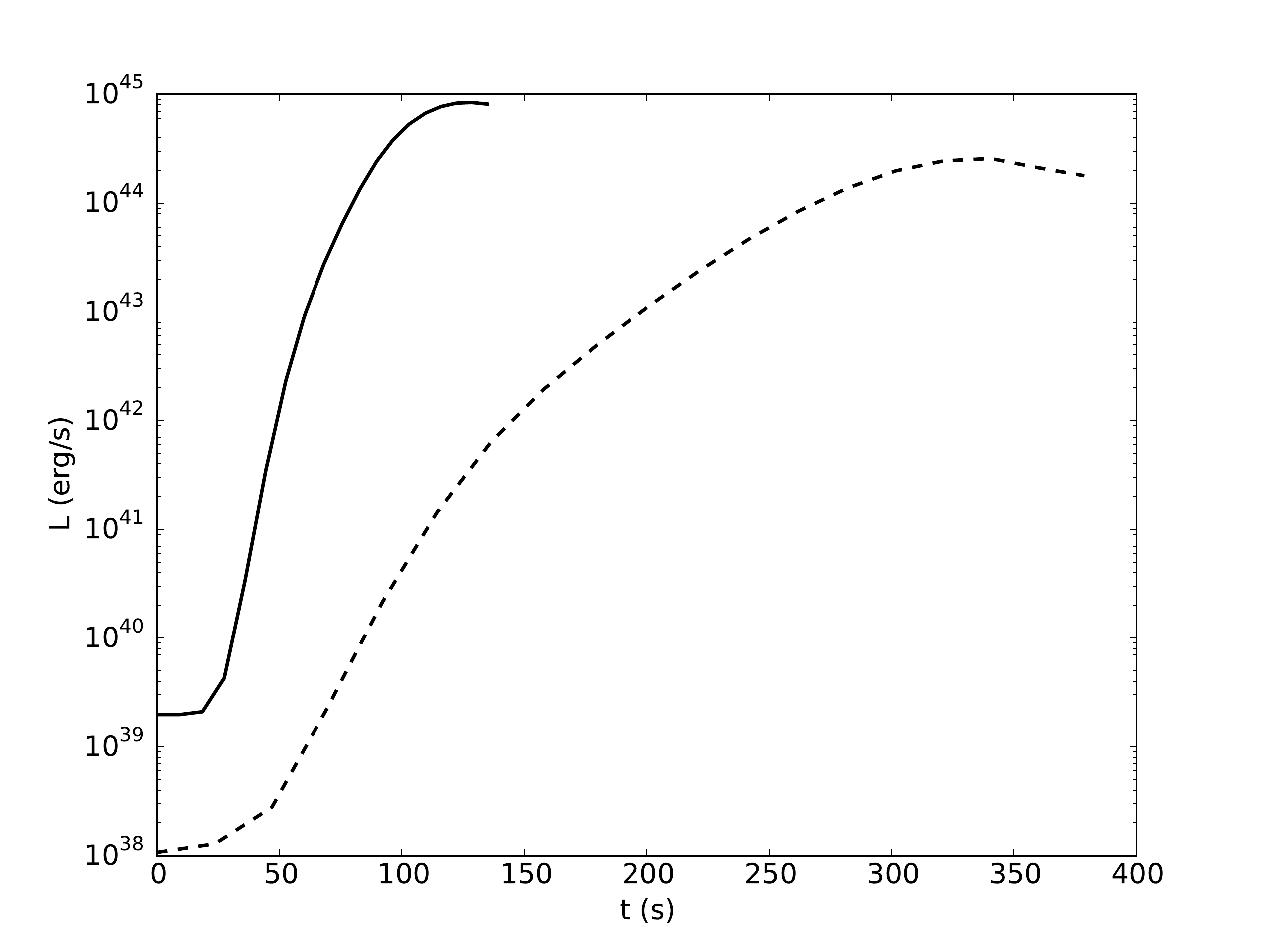}
\caption{ \lFig{87a_atm_lc} Light curves for SN 1987A breakout
  corresponding to the 2 different stellar atmospheres in
  \Fig{87a_atm}, a power law fit to the initial KEPLER model of the
  form $\rho = \alpha_1\, \mathrm{log}_{10} \, r + \beta_1$ (solid)
  and a shallower gradient fit of the form $ \rho = (\alpha_1/2)\,
  \mathrm{log}_{10} \, r + \beta_1$ (dashed). Despite beginning with
  the same shock wave, breakout through the two atmospheres is
  significantly different. The shallower atmosphere produces a longer
  and dimmer breakout than both the steeper model and comparison
  results from other simulations of this event. Thus differences in
  the atmospheric gradient can produce corresponding differences in
  the results, and the atmosphere must be treated with
  care.}
\end{figure}

An alternative to a simple power-law extrapolation of the density
structure in the KEPLER presupernova star's outer zones is to use a
model stellar atmosphere.  For the 15\Msun red supergiant models,
realistic model atmospheres were available from the MARCS database
\citep{marcs}. The MARCS atmospheres were calculated using a
specialized code designed to help observers fit spectral data for red
supergiants. The atmospheres include LTE and NLTE effects not
simulated in either KEPLER or CASTRO, as well as the effects of line
blanketing. The 15 \Msun progenitor has \Teff $\sim$ 3550 K, solar
metallicity, and log surface gravity of -0.32. The two MARCS
atmospheres closest to the 15 \Msun\ progenitor were selected based
upon \Teff, metallicity, and log specific surface gravity, then
interpolated to fit the KEPLER progenitor's properties. The MARCS
atmosphere data extend approximately $5\E{12}$ cm below and $1\E{12}$
cm above the photosphere. In order to ensure a smooth transition a
power law is fit to the combined MARCS atmosphere and KEPLER model and
used to replace all RSG15 atmospheres. This replacement provided a
much more accurate atmospheric gradient. The final progenitor
atmospheres are shown in \Fig{rsg15_marcs}.

The atmosphere of a 25 \Msun red supergiant is more complex and has
been less well-studied in simulation. This progenitor's envelope is
very extended ($\sim 10^{14}$ cm) and loosely bound. Once the star
reaches this stage of its life it will have likely lost part or all of
the envelope to winds or other instabilities; uncertainties in mass
loss make it difficult to reliably estimate the atmosphere's final
structure. The section of parameter space explored by MARCS did not
extend near enough to our progenitor star to reliably extrapolate the
data. Thus for RSG25 a power law of the same form as the 87A fit
described above was fit to the existing KEPLER model and
used to replace the progenitor atmospheres.
%
%

\begin{figure}
\includegraphics[scale=0.4, clip=true, trim = 25 0 0 0]{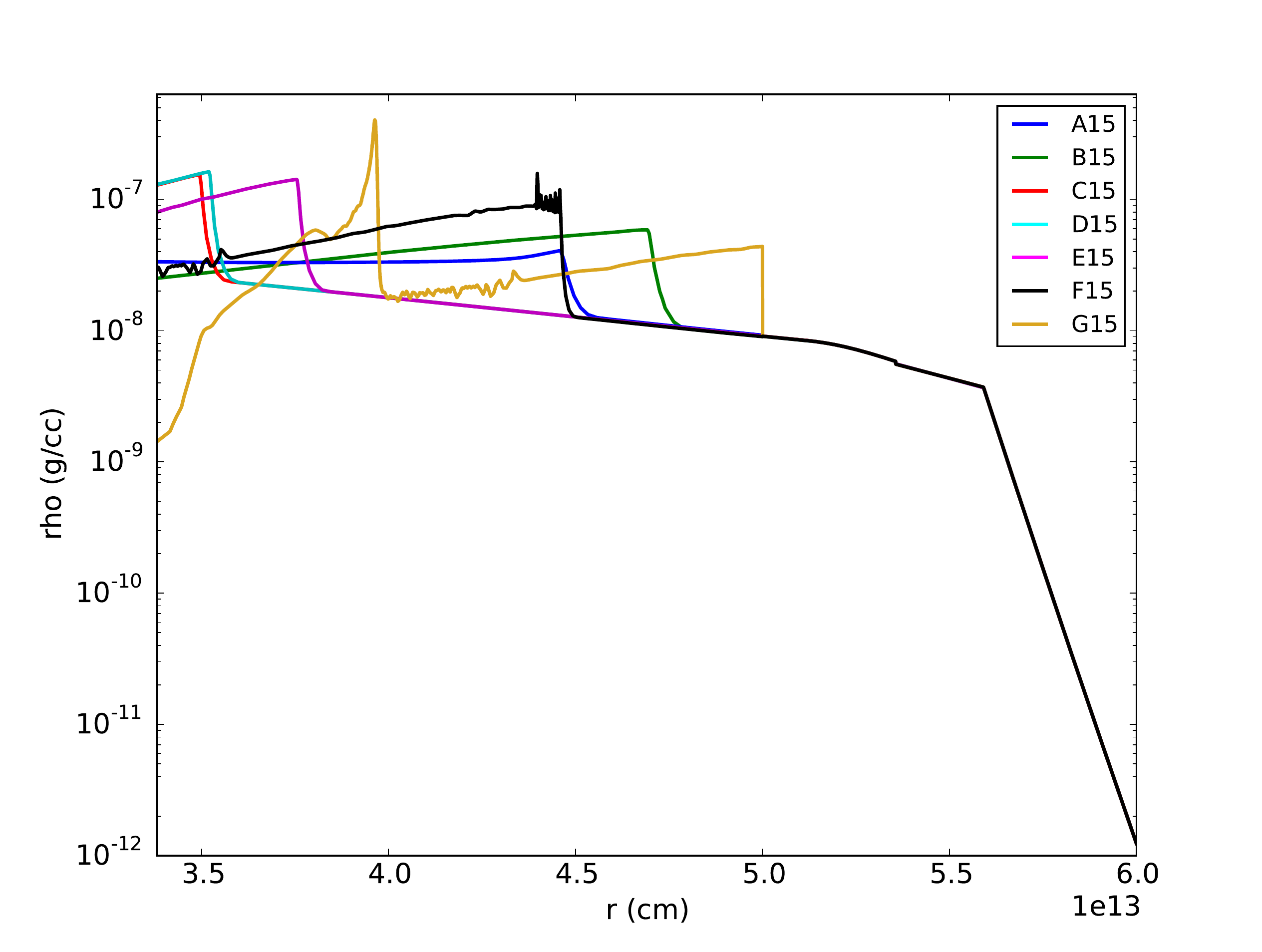}
\caption{ \lFig{rsg15_marcs}Density profiles for the 7 RSG15
  presupernova models from \Fig{rsg15_rhoT}, revised with MARCS
  model atmosphere in place of original KEPLER atmosphere.}
\end{figure}

\subsection{Equation of State and Network}
\lSect{EOS}

The CASTRO simulations used an 18-species network, terminating at nickel, with two auxiliary variables, electron fraction $Y_e$ and mean molecular weight $1/\mu$. Nuclear reactions were turned off for the present study and all species were passively advected. In the regions of interest the composition was dominated by hydrogen and helium at relatively low temperatures, so an ideal gas law plus radiation provided an accurate equation of state. KEPLER has its own general equation of state that is good under all conditions except extremely high density. There was good agreement between the KEPLER pressure, internal energy, temperature and density and the equivalent quantities in CASTRO after the remap.

\subsection{Opacity}
\lSect{opacity}

As discussed in \Sect{opac_theory}, shock breakout is a transition of
the shock wave from an optically-thick to optically-thin region, and
an accurate treatment of opacity is key to obtaining realistic
results. The physics of shock breakout occur at and behind the shock
front where the gas temperature is in the range 4.5\E{4} - 1\E{6} K
when breakout occurs, depending on the model's energy. At these
temperatures and the relevant mass fractions the primary species of
interest are H and He. A Saha solver is used to calculate the
ionization fraction of H and He. The electron abundance is then used
to calculate both a scattering opacity \chiR and a small contribution
to absorptive opacity $\kP_c$, computed by assuming a fixed ratio
$\epsilon = \kP_c/\chiR$\!. Both Thomson scattering and its
contribution to absorptive opacity are grey i.e. insensitive to
frequency. Inverse bremsstrahlung absorption is calculated according
to the equations discussed in \Sect{freefree}. Photoionization
absorption contributed by metals is calculated according to the
equations discussed in \Sect{boundfree}. Line opacities,
bremsstrahlung processes below 4.5\E{4} K, and photoionization
processes below 4.5\E{4} K are not accounted-for, as the former
requires much more detailed calculations (\Sect{boundbound}) and the
grey opacity laws for the latter two processes break down once their
assumptions of complete H and He ionization are significantly
violated. For details on both the theory and implementation of these
opacities, see \Sect{opac_theory}.

CASTRO is an Eulerian code and thus requires a certain amount of mass
in every cell on the grid, or severe instabilities will result. A very
thin ambient medium is therefore placed around the progenitor star to
keep it stable while the processes of interest run. In the case of
shock breakout this ambient medium must be made optically thin to
ensure that its presence does not distort the resulting
lightcurves. The ambient medium is generally made as cold and thin as
possible while still maintaining numerical stability, to avoid
affecting simulation results. If the opacity does not have the correct
low-temperature behavior, this medium can become opaque and influence
the results in an unphysical way. The H-He Saha solver becomes (and
stays) correctly transparent in the ambient medium. Photoionization
opacity is neglected (see previous section). If it were not, this
medium might have some slight absorptive opacity. The medium used in
this study had a density of order $10^{-12}$ g cm$^{-3}$. Most of the
shock breakout transients themselves, and thus our simulations, are
completed by the time the hydrodynamic shock itself reaches the
ambient medium, and if the medium is appropriately transparent, it
will not interact with the radiation either. Our assumed density for
the medium therefore does not affect the breakout transient
hydrodynamically in our simulations.

In reality stars near the ends of their lives -- especially red
supergiants -- are prone to shedding large and uncertain amounts of
mass. Narrow-line Type II supernovae demonstrate clear signs of CSM
interactions. Modeling the full scope of breakout-CSM interactions,
however, is beyond the scope of this paper.


\section{Light Curves and Spectra}
\lSect{lcs}

Light curves and spectra were evaluated by sampling the flux in a
single distant cell, representing the observer. The measured flux is
corrected from the comoving frame back to the lab frame, and then the
entire light curve corrected for light travel time. Because of the
star's curvature, a distant observer will not see the breakout front
erupting uniformly across the star; rather, more distant portions of
the disk will light up at later times since the light must travel
slightly farther to reach the observer. The comparatively small
(3\E{12}\cm) SN1987A progenitor has a light travel time of only 100
seconds, but RSG15 ($\sim$6\E{13}\cm) and RSG25 ($\sim$1\E{14}\cm)
have light travel times of 2000 and nearly 10000 seconds,
respectively. The overall effect of this correction is to smear the
light curve out in time, increasing the peak width and lowering the
peak brightness. We use the same simple light travel correction
formula as T12:
\begin{align*}
L_{obs}(t) &= 2\int^1_0 L(t-\tau) x\, dx,      & \tau = (R_p/c)(1-x)
\end{align*}
This formula makes three assumptions: that the distance to the
observer is large, that the radiation is isotropic, and that the
photospheric radius $R_p$ remains stationary. The first two
assumptions are easily satisfied. The third is less accurate at later
times as the envelope begins to expand, but the speed of the
photosphere is much smaller than the speed of light and $R_p$ can be
effectively taken as constant during the light crossing time.

Spectra are calculated by sampling the individual group fluxes in the
same cell as the light curve. All multi-group models in this paper
were calculated using 64 logarithmically-spaced groups. CASTRO
automatically places any energy at frequencies below the the lower
limit in the lowest group, and any at frequencies higher than the
upper limit in the highest group; thus a failure to resolve the
correct spectral range can be detected by checking for anomalously
high energies in the lowest or highest group. \Tcol is calculated by
selecting the frequency group with the highest energy density and
applying Wien's Law to its central frequency. \Teff is calculated from
the bolometric radiation flux using the standard $F=\sigma_{SB}
\Teff^4$ blackbody relation. We considered breakout to be complete
when the bolometric light curve had declined to at least half of peak
brightness.

\section{Breakout in Red Supergiants}
\lSect{breakoutrsg}
\subsection{Bolometric Light Curve}
Due to their large radii, breakout transients in red supergiants have a much longer time scale than SN 1987A, further lengthened by the low energies considered. Typical durations of the breakout peak are hours to days. For some of the lower energies, the supernova fails almost completely. For others the envelope is ejected, but not the helium core. The wide range of kinetic energies examined results in a diverse set of peak luminosities. Perhaps unfortunate for their detection, duration and peak brightness are inversely related - that is, the brighter the breakout, the shorter it will likely be. 

In RSG15, the final kinetic energies for the VLE models ranged from $6.6\E{46}$ erg to $1.23\E{50}$ erg as listed in \Tab{rsgresults}. Their peak luminosities ranged from 9.57\E{39} - 2.02\E{44} erg. The two standard-energy models, F15 (0.5 B) and G15 (1.2 B), had peak luminosities of 8.07\E{44} and 1.66\E{45} respectively. Bolometric light curves for RSG15 are shown in \Fig{rsg15_lc}. Full-width half-maximum durations range from 3 - 35 h, with an outlying 70 h for the lowest energy model, A15. Though mass loss leaves RSG15 and RSG25 with similar presupernova masses (12.79 \Msun vs. 15.84 \Msun\!), the radius of RSG25 is nearly double the radius of RSG15. Breakouts with similar energies are thus expected to have longer light curves and lower peak energies in RSG25 versus RSG15. Bolometric light curves for RSG25 (\Fig{rsg25_lc}) show peak luminosities around $10^{42}$ erg/s and durations in the 25 - 70 h range, excepting E25, whose kinetic energy at breakout was significantly higher. 

Models B15, C15, E15, and G15 form a sequence where each model's final kinetic energy increases by approximately an order of magnitude, spanning $10^{48} - 10^{51}$ erg. Their peak luminosities span five orders of magnitude ($10^{41} - 10^{45}$ erg/s), with the largest relative increases occurring between B15 and C15, and C15 and E15, about a factor of 20. The duration of B15 (1.54\E{48}) is about 35 hours while the durations of C15 (1.12\E{49} erg), E15 (1.23\E{50} erg), and G15 (1.2\E{51} erg) are 8.1h, 3.1h, and 1.4h, respectively, meaning a significant shift in breakout duration occurs between $\sim 10^{48}$ and $10^{49}$ erg of final kinetic energy.

Models D25 and C15 have similar kinetic energies at breakout (3.52\E{48} vs. 4.98\E{48} erg); their durations, however, are significantly different. C15 has a duration of 8.1 hours vs. 25.9 for D25. Models E25 and D15 have similar energies (1.10\E{49} vs. 2.13\E{49} erg) and similar durations of 9.3 vs. 5.3 hours. Again a strong shift in behavior is observed with energy, in this case the influence of progenitor structure on breakout. The much larger radius of RSG25 means that a shock will spend much more time traveling through the envelope than it would in RSG15. This longer travel time and increased interaction with the envelope may amplify the effect of progenitor structure on the resulting breakout. Shock waves energetic enough to traverse the envelope relatively quickly may have less time to interact and therefore show less variation with progenitor structure. Kinetic energy therefore emerges again as the most important parameter governing breakout behavior. This agrees with the analytic equations derived by \citet{piro}, discussed further in \Sect{analytic}, which give a dependence $L \propto E_{48}^{1.36}$.


\begin{figure}
\includegraphics[scale=0.4, clip=true, trim = 25 0 0 0]{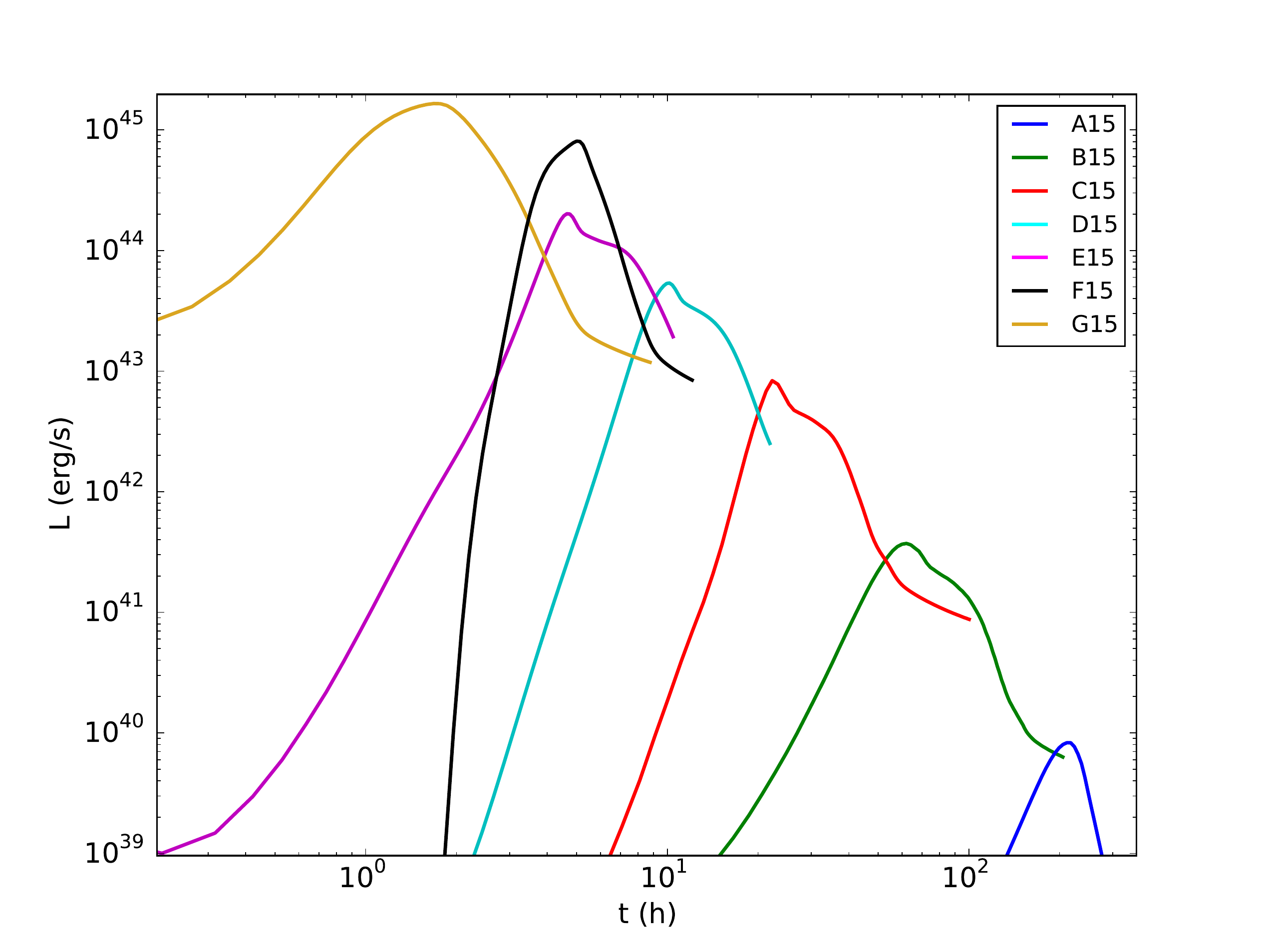}
\caption{ \lFig{rsg15_lc} Bolometric light curves for RSG15 shock breakouts at 7 different final kinetic energies ranging from $6.58\E{46}$ (A15) to $1.20\E{51}$ (G15), calculated by CASTRO. Both peak luminosity and breakout flash duration show clear and significant variations with explosion energy. Light curves are shown with a logarithmic time axis due to the extreme variation in duration with energy. The slight anomalies in the light curve post-peak are discussed in \Sect{breakoutrsg}.}
\end{figure} 

\begin{figure}
\includegraphics[scale=0.4, clip=true, trim = 25 0 0 0]{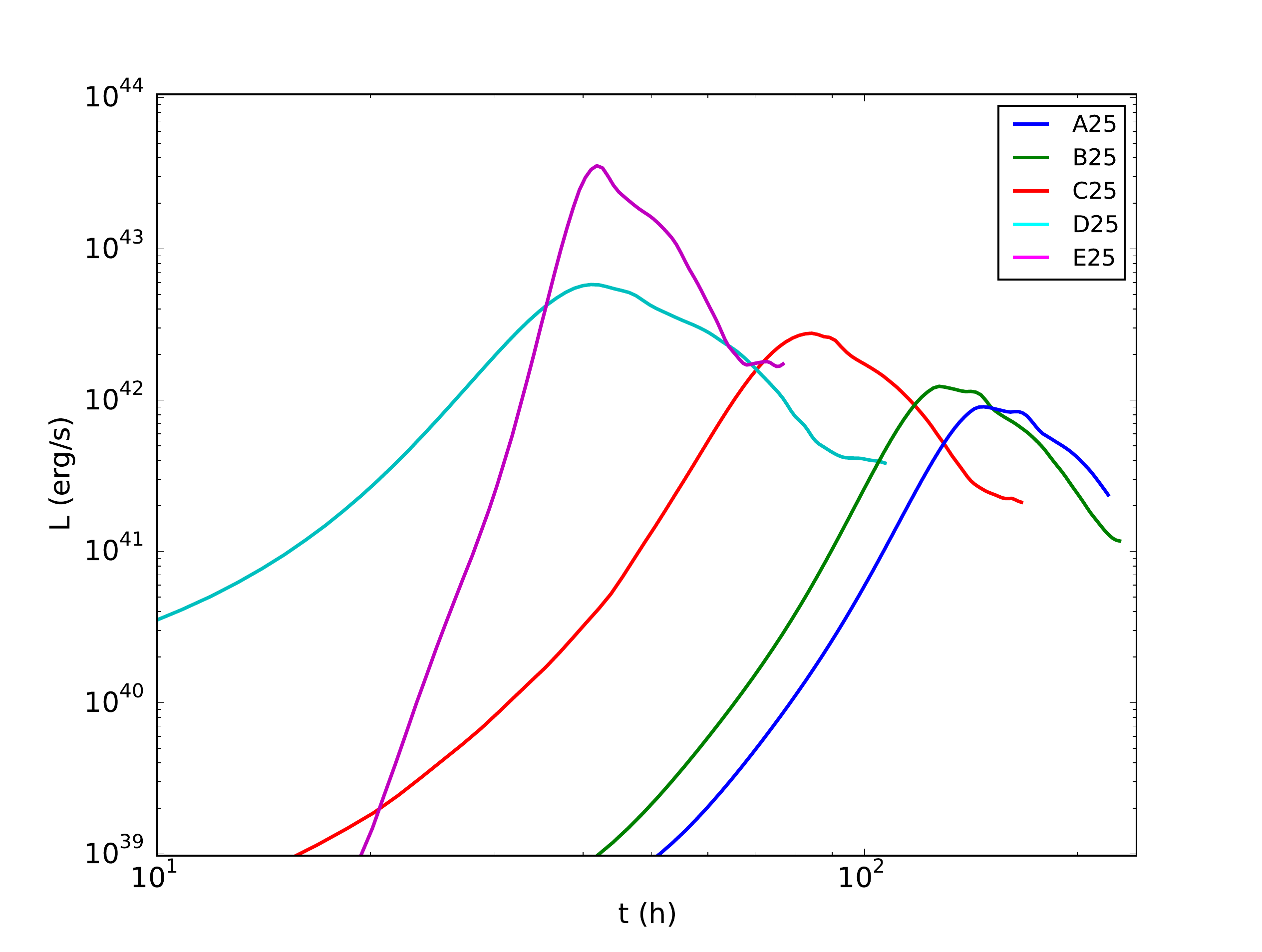}
\caption{ \lFig{rsg25_lc} Bolometric light curves for RSG25 shock breakouts at 5 different explosion energies ranging from $1.38\E{48}$ (A25) to $1.10\E{49}$ (E25), calculated by CASTRO. Light curves are shown with a logarithmic time axis due to the extreme variation in duration with energy.}
\end{figure} 

Both RSG15 and RSG25 show minor anomalies in their light curves post-peak. RSG15's shock breakouts show a curious blip in their in decline rate post-peak - initially declining quite steeply, but briefly changing to a shallower slope. Four of RSG15's breakouts display this anomaly and it is correspondingly extended in the longer-duration light curves. RSG25 shows a similar variation post-peak, although in this case it shows multiple peaks/changes in the rate of decline. None of these small features represent significant variations relative to the light curve of each breakout, but the fact that they appear in all light curves of a single progenitor merits further investigation. It is unclear what physical behavior causes these anomalies, but their appearance seems to be related to the formation of a high-density spike at the photosphere as the hydrodynamic shock begins to expand the envelope into the ambient medium. This spike is likely an artifact of 1D simulation. As noted, none of these features are significant relative to the overall light curves, so they can be neglected.

\begin{figure}
\includegraphics[scale=0.4, clip=true, trim = 25 0 0 0]{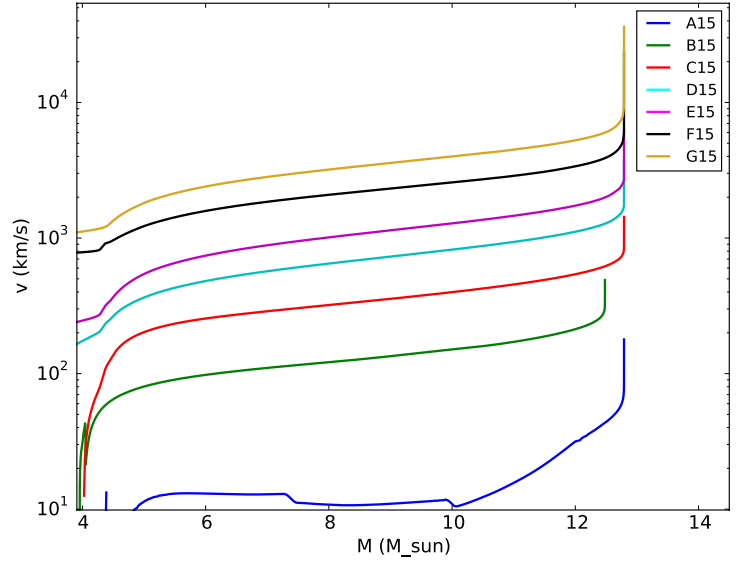}
\caption{ \lFig{kep_vel} Late-time velocity profiles as calculated by KEPLER for RSG15 models. {\red Model B15 and only B15 was launched by the abrupt removal of 0.2 \Msun \ of neutrino mass loss from its iron core, hence the different end point on the plot.}}
\end{figure} 

{\red Only a small fraction of the envelope achieves escape speed in
  Model A15, as shown in \Fig{kep_vel}, and the dynamics of this
  model are not as well determined as the others.}

\subsection{Color Temperature}

{\red CASTRO's multigroup transport was used to generate the spectra
  of RSG15 breakouts with various shock energies. These are shown in
  \Fig{rsg15_spec}. All have a dilute blackbody form.
Models A15 and B15 were excluded from the multigroup simulation
because their large optical depth at breakout posed computational
difficulty and the opacity was not realistic anyway (see below).  All
multigroup models were calculated using 64 logarithmically-spaced
groups. Model {\red C15} was run with a frequency range $1.5\E{14} -
5\E{16}$ Hz. Models D15 and E15 were run with the range 5.0\E{14} -
1.0\E{17} Hz. Models F15 and G15 were run with range 1.5\E{14} -
1.0\E{17} Hz.

\begin{figure}
\begin{center}
\includegraphics[scale=0.4, clip=true, trim = 25 10 70 50]{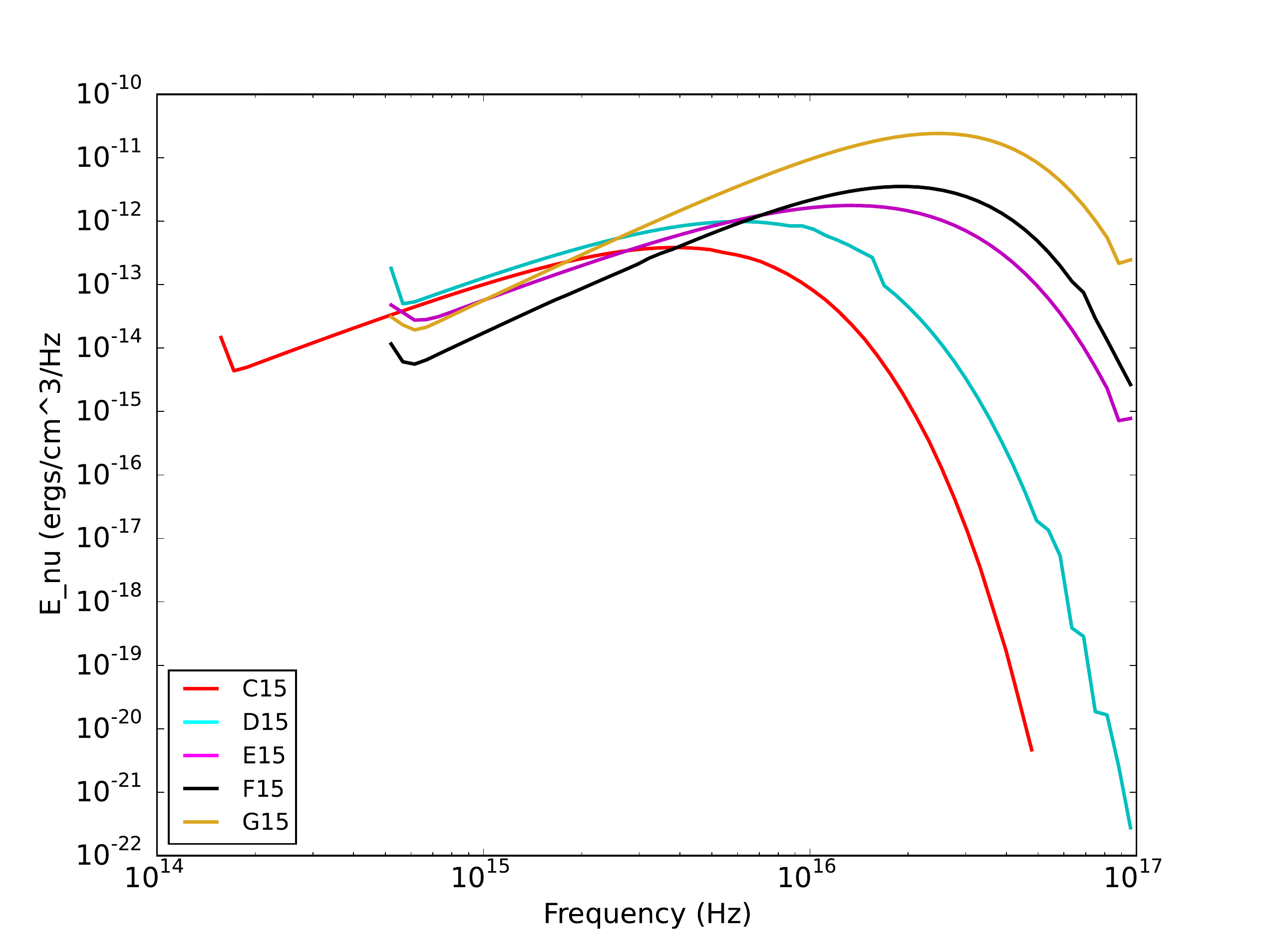}
\caption{ \lFig{rsg15_spec} Multigroup spectra of RSG15 shock breakouts sampled at peak color temperature for 5 different explosion energies, calculated by CASTRO. These spectra peak largely in the hard UV ($>10^{16}$ Hz), but still have significant output in the visible and IR bands.}
\end{center}
\end{figure} 

Near peak luminosity in Models C15 - E15 \Tcol ranges from $6.4\E{4}$
to $3.18 \E{5}$ K, moving the blackbody emission peak down from the
soft X-ray bands into the hard UV.  Previous works have found a color
temperature higher than the effective temperature by a factor of
roughly 2 - 3 \citep{eb92, t12, kc78}. Ratios of \Tcol to $T_{ef}$ in
Models C15 - G15 in \Tab{rsgresults} range from 1.4 to 2.2, similar to
the values suggested by {\red\citet{Rab10}}.

However, as noted in \Sect{vle_tcol}, these previous works considered
only high energy explosions where electron scattering grossly
dominated the opacity. In the low-energy breakouts here, a
larger absorptive component helps to thermalize the diffusing photons.
Given the small fraction of absorptive opacity needed for
thermalization (\Sect{vle_tcol}) and the opacities shown in Figs. 1 -
3, we expect the near equality of color and effective temperature at
peak for Models A15, B15, and C15. Higher energies, approaching those
of SN 1987A, should be dominated by electron scattering with \Tcol
close to the maximum values given in \Tab{rsgresults} (Models F15 and
G15). Models D15 and E15 should have \Tcol somewhere between the
electron scattering value and $T_{ef}$.  This is actually good news
for the low energy cases since it implies that breakout there can be
calculated without recourse to multigroup transport, i.e., by codes
using a single temperature.

Further work on these very low energy cases would benefit from using
an atomic opacity database such as OPAL, but the implementation of
this poses difficulty. OPAL tables give the total opacity and do not
distinguish the absorptive and scattering components, as is needed for
CASTRO. In the past EB92 overcame a similar limitation by assuming
scattering opacity to come entirely from Compton scattering, computing
this value, then subtracting it from the OPAL value to get an
absorptive opacity. In cases where the scattering opacity is dominant,
this technique can lead to errors as one large number is being
subtracted from another similar number, giving an uncertain small
difference.}

\begin{deluxetable*}{ccccccccccc}
\tablewidth{0pt}
\tablecaption{VLE Breakout Model Results}
\tablehead{
\colhead{Model} &
\colhead{Star} &
\colhead{$v$\tablenotemark{a} (km/s)} &
\colhead{KE$_b$\tablenotemark{b} (ergs)} &
\colhead{KE$_f$\tablenotemark{c} (ergs)} &
\colhead{$L_{p}$ (erg/s)} &
\colhead{$L_{cr}$\tablenotemark{d} (erg/s)} &
\colhead{$\Delta t$\tablenotemark{e}  (h)} &
\colhead{Max \Teff (K)} &
\colhead{Max \Tcol (K)} &
\colhead{$\lambda_{max}$ (\AA)}}
\startdata
A15 & RSG15 & 80 & 3.86\E{46} & 6.58\E{46} & 9.50\E{39} & 9.57\E{39} & 68.4 & 8.15\E{3} & \nodata & \nodata\\
B15 & RSG15 & 150 & 6.43\E{47} & 1.54\E{48} & 3.89\E{41} & 3.82\E{41} & 35.2 & 2.06\E{4}  & \nodata & \nodata\\
C15 & RSG15 & 320 & 4.98\E{48} & 1.21\E{49} & 8.39\E{42} & 8.33\E{42} & 8.1 & 4.44\E{4} & 6.40\E{4} & 797\\
D15 & RSG15 & 650 & 2.13\E{49} & 5.04\E{49} & 5.43\E{43} & 5.38\E{43} & 5.3 & 7.08\E{4} & 1.15\E{5} & 443\\
E15 & RSG15 & 1200 & 5.43\E{49} & 1.23\E{50} & 2.13\E{44} & 2.02\E{44} & 3.1 & 9.97\E{4} &  2.24\E{5} & 228\\
F15 & RSG15 & 1920 & 2.16\E{50} & 5.07\E{50} & 8.25\E{44} & 8.07\E{44} & 1.83 & 1.40\E{5} & 2.35\E{5} & 217\\
G15 & RSG15 & 2400 & 2.54\E{50} & 1.20\E{51} & 1.68\E{45} &1.66\E{45} & 1.42 & 1.67\E{5} & 3.18\E{5} & 160\\
A25 & RSG25 & 150 & 1.38\E{48} & \nodata & 9.07\E{41} & 9.03\E{41} & 67.0 & 1.57\E{4} & \nodata & \nodata\\
B25 & RSG25 & 179 & 1.53\E{48} & \nodata & 1.23\E{42} & 1.23\E{42} & 57.9 & 1.70\E{4} & \nodata & \nodata\\
C25 & RSG25 & 210  & 2.27\E{48} & 6.12\E{48} & 2.76\E{42} & 2.77\E{42} & 37.0 & 2.08\E{4} & \nodata & \nodata\\
D25 & RSG25 & 280 & 3.52\E{48} & \nodata & 5.85\E{42} & 5.83\E{42} & 25.9 & 2.51\E{4} & \nodata & \nodata\\
E25 & RSG25 & 480 & 1.10\E{49} & \nodata & 3.67\E{43} & 3.61\E{43} & 9.3 & 3.97\E{4} & \nodata & \nodata\\
\enddata
\tablenotetext{a}{Velocity at breakout.}
\tablenotetext{b}{Kinetic energy at breakout.}
\tablenotetext{c}{Final kinetic energy of supernova as calculated by KEPLER.}
\tablenotetext{d}{Corrected for light travel time.}
\tablenotetext{e}{Full-width half-max of travel-time-corrected light curve.}
\lTab{rsgresults}
\end{deluxetable*}

\begin{deluxetable*}{cccccccc}
\tablewidth{0pt}
\tablecaption{Comparison to Other Models}
\tablehead{
\colhead{Model} &
\colhead{KE$_f$\tablenotemark{a} (ergs)} &
\colhead{$L_{peak}$ (erg/s)} &
\colhead{$L_{\mathrm{K}}$\tablenotemark{b} (erg/s)} &
\colhead{Pred. $L$ (erg/s)} &
\colhead{Max \Teff(K)} &
\colhead{ \Teff$\!_\mathrm{,K}$\tablenotemark{c}(K)} &
\colhead{Pred. \Teff(K)}}
\startdata
A15 & 6.58\E{46} & 9.50\E{39} &  5.09\E{39} & 4.25\E{39} & 8.15\E{3} & 6.93\E{3} & 6.29\E{3}\\
B15 & 1.54\E{48} & 3.89\E{41} &  3.94\E{41} & 3.10\E{41} & 2.06\E{4}  & 1.99\E{4}& 1.84\E{4}\\
C15 & 1.21\E{49} & 8.31\E{42} & 6.32\E{42} & 5.11\E{42} &  4.44\E{4} & 4.02\E{4} & 3.71\E{4}\\
D15 & 5.04\E{49} & 5.43\E{43} &  3.95\E{43} & 3.56\E{43} & 7.08\E{4} & 6.36\E{4} & 6.02\E{4}\\
E15 & 1.23\E{50} & 2.13\E{44} &  1.17\E{44} & 1.20\E{44} & 9.97\E{4} & 8.34\E{4} &  8.15\E{4}\\
F15 & 5.07\E{50} & 8.25\E{44} & 6.17\E{44} & 8.06\E{44} & 1.40\E{5} & 1.30\E{5} & 1.31\E{5}\\
G15 & 1.20\E{51} & 1.68\E{45} & 1.76\E{45} & 2.65\E{45} & 1.67\E{5} & 1.69\E{5} & 1.77\E{5}\\
\enddata
\tablenotetext{a}{Final kinetic energy as measured in KEPLER}
\tablenotetext{b}{Peak luminosity of light curve in KEPLER.}
\tablenotetext{c}{Peak effective temperature in KEPLER.}
\lTab{comparison}
\end{deluxetable*}

\subsection{Comparison to KEPLER Results}
\lSect{kepler_comp} 
\begin{figure}
\includegraphics[scale=0.4, clip = true, trim = 25 0 0 0]{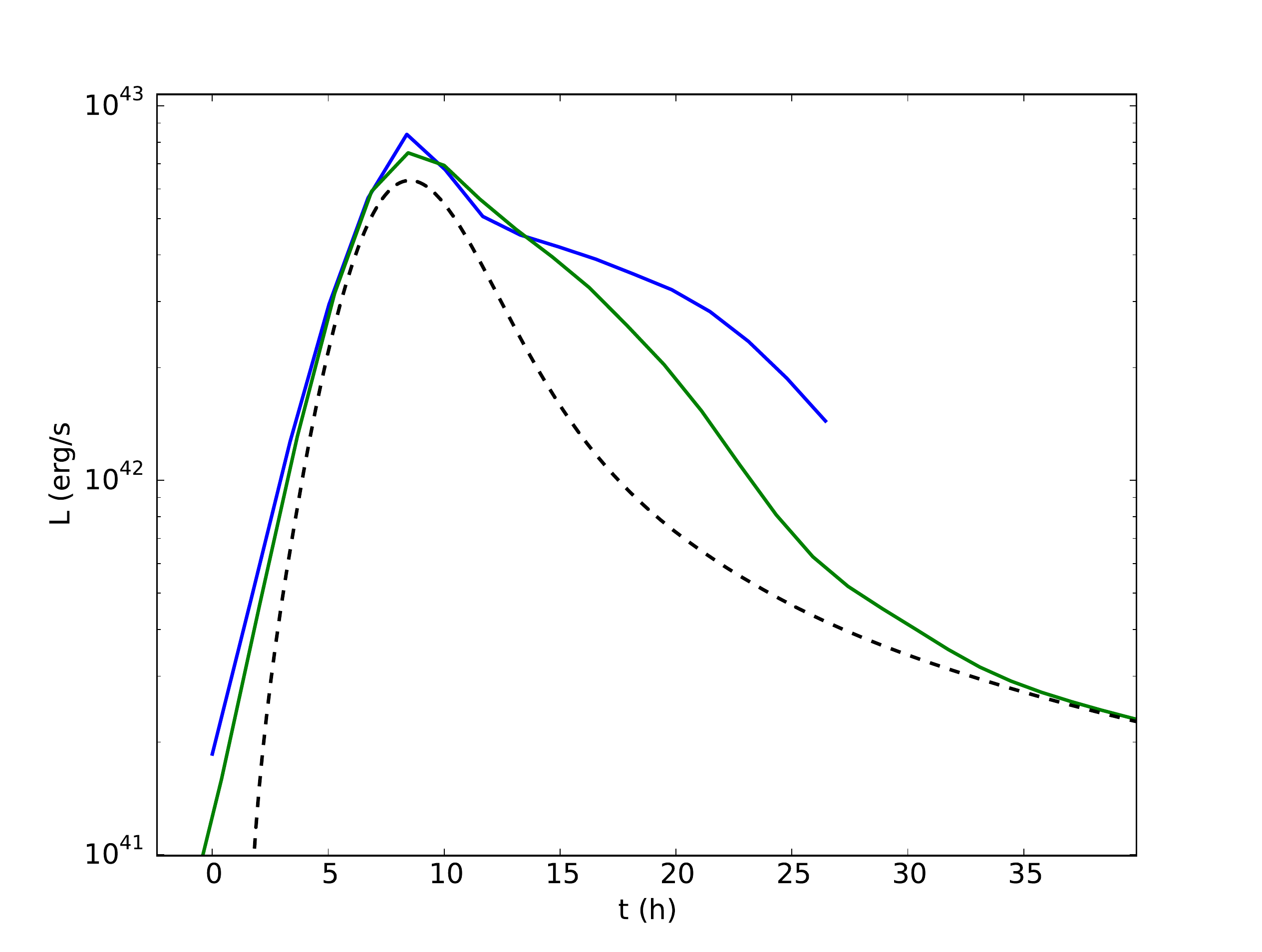}
\caption{\lFig{c15_atm}Light curves for breakout in RSG15, model C15, calculated in CASTRO for 2 different stellar atmospheres: MARCS fit (blue); and fit to initial KEPLER data (green). The dashed line shows the light curve for the same model as computed entirely in KEPLER with fine zoning.}
\end{figure}

As noted in Section \Sect{progenitor}, the model atmospheres were
replaced in the move from KEPLER to CASTRO. \Fig{c15_atm} shows
bolometric luminosity of two breakouts calculated in CASTRO, one using
the MARCS atmosphere (blue) and one using a fitted version of the
original KEPLER atmosphere (green). The differences are slight but
present. KEPLER itself calculates a bolometric light curve that
differs from CASTRO, but the divergence appears to be
resolution-related, as finer zoning in KEPLER significantly reduced
the difference. Full results are shown in \Tab{comparison}. KEPLER
slightly underestimates the peak luminosity of the CASTRO results and
falls within a factor of 1 - 3 of the analytic predictions. KEPLER can
therefore produce a useful estimate of the bolometric light curve,
even with its simpler radiation transport. Of the two codes, CASTRO is
the only one that can compute a color temperature, so that result
cannot be compared directly. However, if the theoretical work
considered in \Sect{opac_theory} can be used to estimate the location
of the chromosphere, a color temperature may still be calculated from
KEPLER results.

\subsection{Comparison to Analytic Predictions}
\lSect{analytic}

Analytic predictions for shock breakout in ``normal'' Type IIp supernovae already exist in the literature \citep{tominagatypeII, matzner99, katz10}. \citet{piro} extended these formulas to consider the specific case of low-energy supernovae. These formulas predict the bolometric luminosity and timescale of breakouts based on the properties of the progenitor star and the explosion and can be tested against our numerical results. \citet{piro} gives:
\begin{align*}
L_{bo} &= 1.4\E{41} \frac{E_{48}^{1.36}}{\kappa_{0.34}^{0.29}M_{10}^{0.65}R_{1000}^{0.42}}\left(\frac{\rho_1}{\rho_*}\right)^{0.194} \mathrm{erg/s}\\
T_{obs} &= 1.4\E{4}\frac{E_{48}^{0.34}}{\kappa_{0.34}^{0.068}M_{10}^{0.16}R_{1000}^{0.61}}\left(\frac{\rho_1}{\rho_*}\right)^{0.049} \mathrm{K}
\end{align*}
where $E_{48} = E_{kin}/10^{48}, M_{10} = M_{ej}/10 \Msun\!, R_{1000} = R_*/1000 R_\odot,$ and $\kappa_{0.34} = \kappa/0.34$. Stellar radii are given in \Tab{models}. It is assumed in both cases that the ejecta mass $M_{ej}$ is equal to the size of the hydrogen envelope, that the adiabatic index $\gamma = 5/3$, and that the factor $(\rho_1/\rho_*)
\sim 1$. The results are shown in Table 4.

There is some ambiguity in the equations as to when the quantity $E_{48}$ should be measured; the kinetic energy at breakout differs from the ejecta's final kinetic energy at infinity. For RSG15, the peak luminosity predictions fell much closer to the KEPLER and CASTRO results when $E_{48}$ was assumed to be kinetic energy at infinity as measured in KEPLER. The analytic predictions and the KEPLER results are very close and both slightly underestimate the CASTRO luminosities. RSG25 shows much greater variance; the analytic predictions underestimate the numerical results by a factor of 4 - 10, with the inaccuracies increasing with kinetic energy. The analytic formulas therefore give reasonable if not precise estimates for a breakout's brightness.

\section{Observing Prospects}

The most realistic and refined simulations are still of little use without observations to test them. This section discusses current and near-future surveys as well as recovered candidates for VLE SNe and CCSNe failures, and provides considerations and guidance for future observations of the transients discussed in this paper, especially
with regard to the expected color temperatures in VLE SNe as opposed to standard-energy breakouts.

\Fig{kep_lc} shows the evolution of the RSG15 models post-shock breakout as simulated by KEPLER. \Fig{kep_rsg25lc} shows the same for the RSG25 models. The RSG15 plateau durations vary significantly with energy, as might be expected. Plateau magnitudes tend to be some 1.5 - 2 orders of magnitude lower than the breakout peak. 

\begin{figure}
\includegraphics[scale=0.4, clip=true, trim = 25 0 0 0]{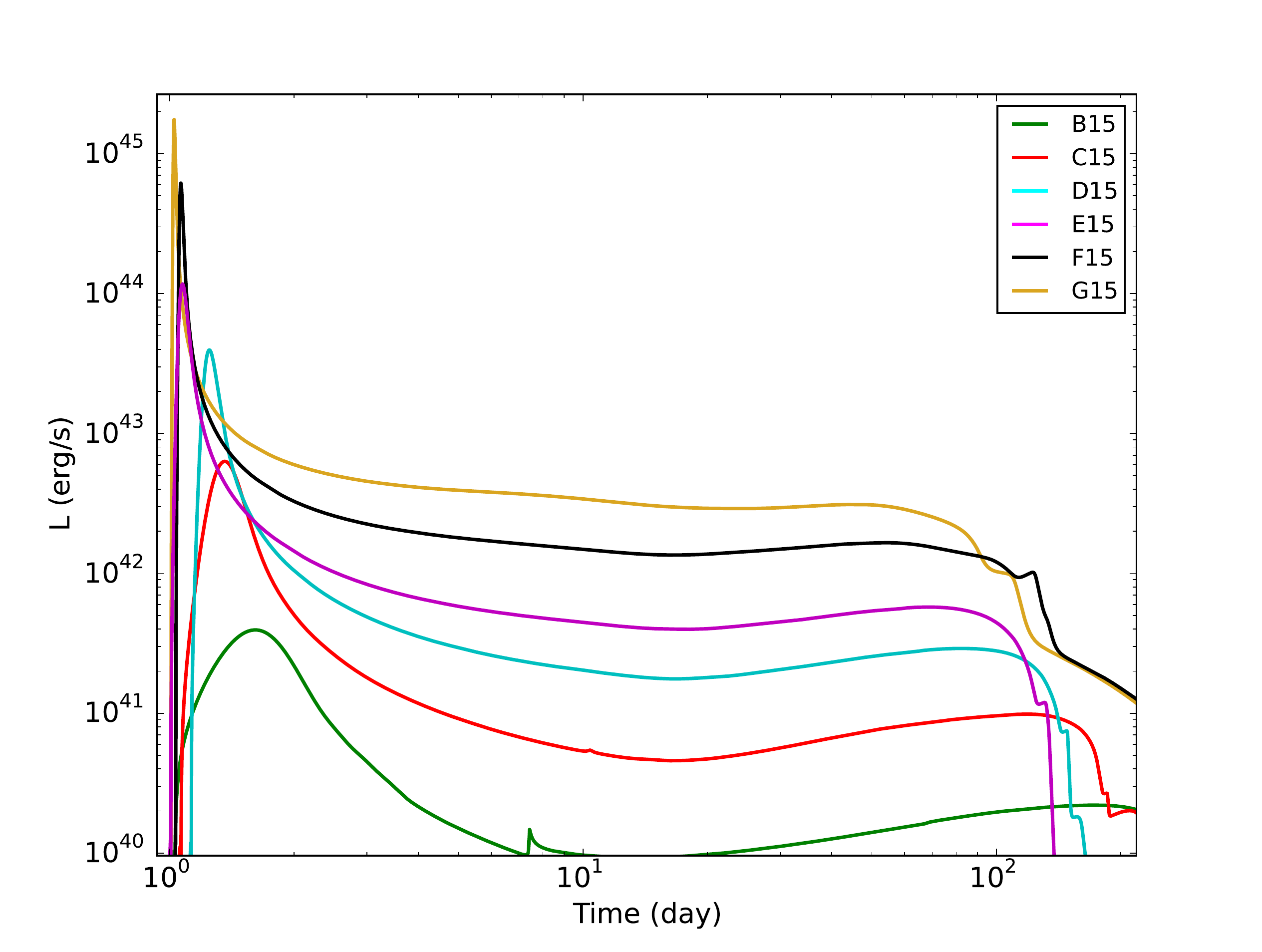}
\caption{ \lFig{kep_lc}Late-time light curves calculated by KEPLER showing the evolution and plateau phase of RSG15 models. Calculations assumed opacity due to electron scattering and an opacity floor of 10$^{-5}$ cm$^{2}$ g$^{-1}$.}
\end{figure} 

\begin{figure}
\includegraphics[scale=0.4, clip=true, trim = 25 0 0 0]{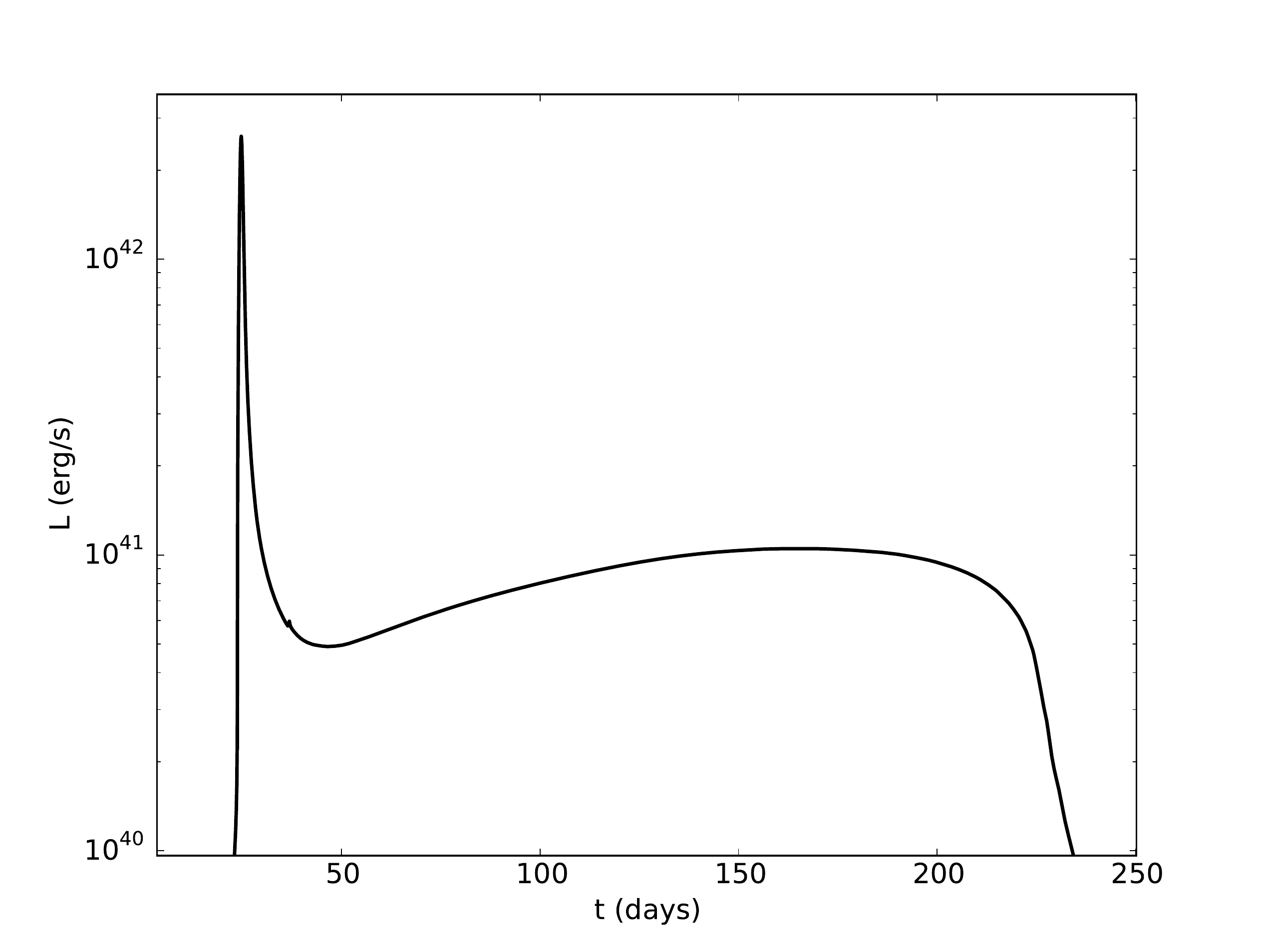}
\caption{ \lFig{kep_rsg25lc}Late-time light curve calculated by KEPLER for the RSG25 {\red model.}}
\end{figure}

\subsection{VLE SNe Breakout in Optical \& IR}
\lSect{vle_ir}

The practical upshot of the opacity effects discussed in \Sect{vle_tcol} is that the color temperatures of these faint breakouts are significantly cooler than those of more energetic events, both because of their lower shock energies and because of the convergence of \Tcol\ and \Teff\!. Although the bolometric luminosity of these events is much lower than normal CCSNe, more of the energy will be emitted at low frequencies, and in an IR band the low-energy breakout may actually appear \textit{brighter} than its more bolometrically-energetic counterpart. The reported detections of shock breakout by the Kepler satellite \citep{kepler2}, which observes primarily in the optical and near-infrared, emphasize the advantage of this cooler spectrum. 

A simple estimate of the brightness of breakout transients in any given band can be found by assuming the spectrum is a blackbody at \Tcol\ and calculating the fraction of blackbody energy inside that bandpass. More accurate calculations would be done using a measured filter curve. The Kepler satellite has an IR bandpass of 0.4 - 0.9 microns. \Fig{rsg15_bandpass} shows the results of calculating the fraction of blackbody energy emitted within that bandpass, assuming that \Tcol is equal to the maximum \Teff\ in low-energy events. Color temperature is unlikely to go lower than this value, meaning these curves represent upper bounds. The dynamic range of peak luminosities is significantly compressed, as the brighter breakouts also have higher color temperatures. However the duration of the transient is unaffected, and thus breakout events of similar luminosity can still be distinguished in energy by measuring the duration. The peak luminosities for these filtered curves range from 4\E{41} - 4\E{39} erg/s. These are still dim events, but not out of reach of current and future surveys; notably, they are significantly brighter for longer than a standard-energy breakout in these bands.

\Fig{ir_comparison} shows simulated KEPLER light curves for the same 0.4 - 0.9 micron range. The KEPLER code cannot directly compute a color temperature, but based on analytic and numerical results we can make some assumptions about its value. The black curve shows the predicted light curve in the 0.4 - 0.9 micron range of an explosion with final KE 2.4 B, calculated by applying formulas from \citet{nakar10} to locate the chromosphere and using the gas temperature at that layer as a color temperature. The red curve shows a similar prediction for a VLE SNe at 1.2\E{49} erg, corresponding to model C15. At this energy color temperature is predicted to converge with effective temperature and a light curve can be made with a 1-T code that assumes \Tcol = \Teff. The low-energy supernova, though nearly a thousand times less energetic, peaks at a higher luminosity and stays bright for longer in the IR. Again, these are still dim events, but not out of reach.

\begin{figure}
\begin{center}
\includegraphics[scale=0.4, clip=true, trim = 25 0 0 0]{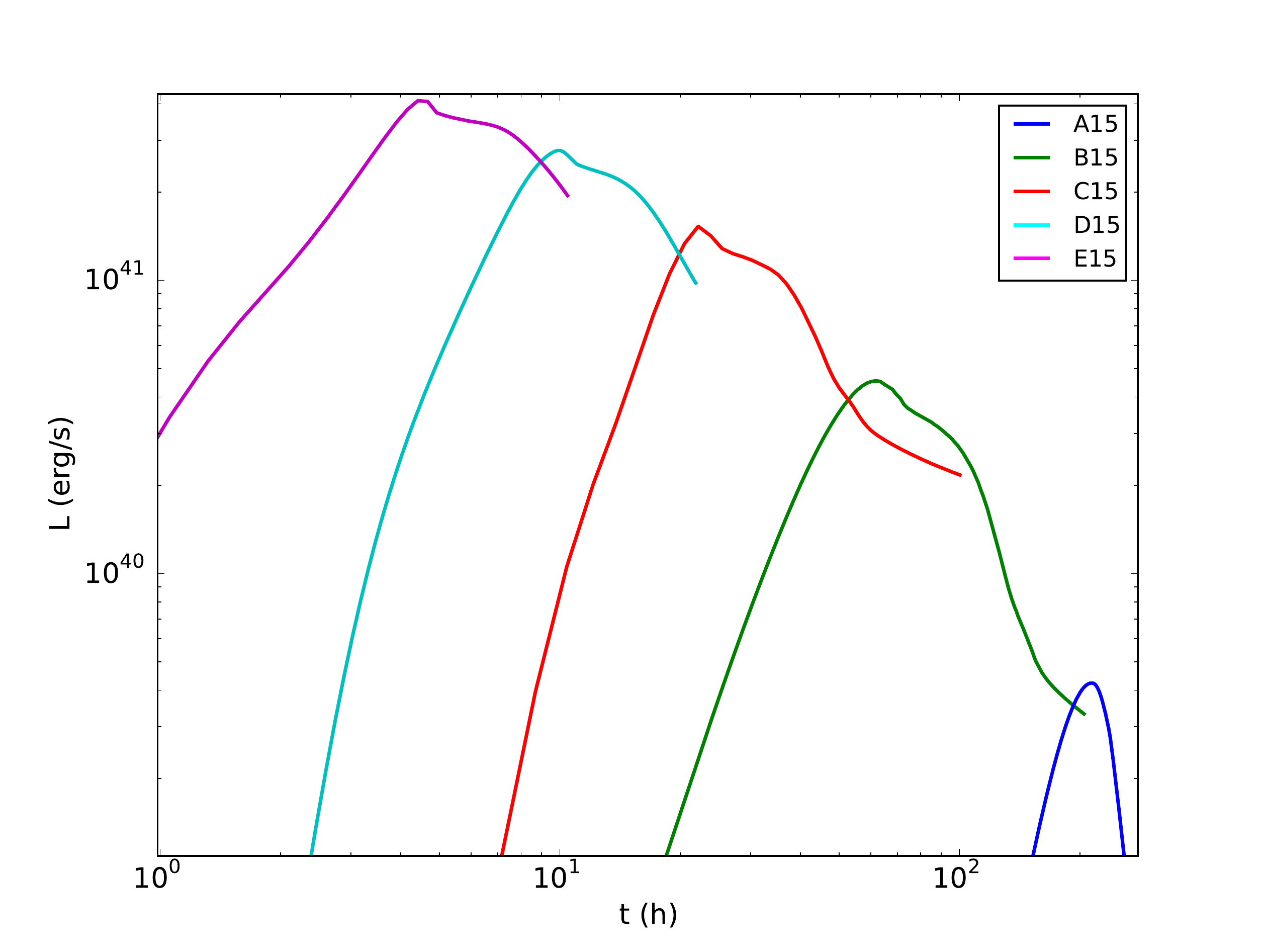}
\caption{ \lFig{rsg15_bandpass} RSG15 models as they would be observed in the band 0.4 - 0.9 microns, assuming color temperature is equal to the maximum effective temperature. As color temperature cannot drop below effective temperature, these light curves therefore represent upper bounds. Higher-energy breakouts have greater bolometric luminosity, but also have higher color temperature, which suppresses their peak luminosity in optical and IR bands. Duration remains unaffected. Models F15 and G15 are not shown here as the color temperature approximation is not applicable at those energies.}
\end{center}
\end{figure}

\begin{figure}
\begin{center}
\includegraphics[scale=0.4, clip=true, trim = 25 0 0 0]{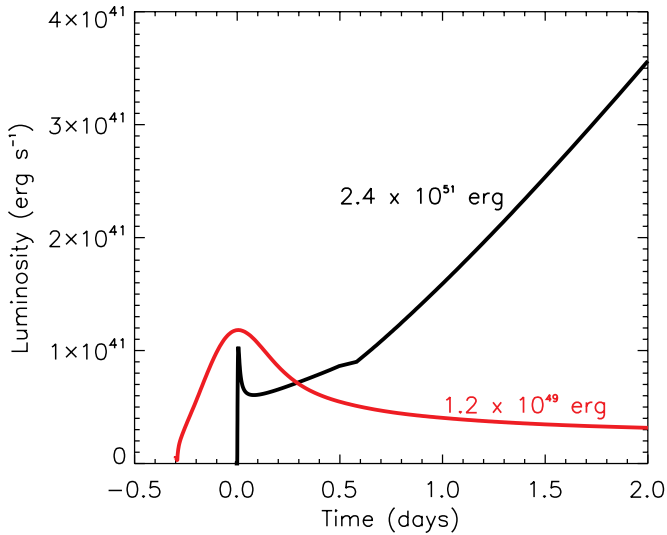}
\caption{\lFig{ir_comparison}Shock breakout light curves simulated in KEPLER and convolved with the bandpass 0.4 - 0.9 microns. The black line represents a 2.4 B explosion, while the red one shows a VLE SNe at only 1.2\E{49} erg (corresponding to model C15). Note that although the VLE SNe is dimmer bolometrically, its breakout flash seen in this band is both brighter and longer than the more energetic event.}
\end{center}
\end{figure}


The cooler temperatures of RSG breakouts make studying VLE SNe at larger distances a more attractive prospect, as redshifting will move the light into optical and IR windows. But the majority of the supernova's energy is still emitted at wavelengths shorter than the Lyman $\alpha$ cutoff ($> 95$\% at a color temperature $1\E{5}$ K), and if an extragalactic supernova is assumed to be embedded in a UV-absorbing ISM within its host galaxy, this energy may be absorbed at the source. Since these transients are already faint, nearby (z $\ll$ 1) events are the primary target, and the majority of absorption will occur either at the source or in the Galaxy. The UV attenuation will therefore depend on the viewing angle through the Galaxy and the unknown circumstellar medium at the source. Nearby supernovae will still offer a better target.

\subsection{Searches for Failed Supernovae}

\citet{kochanekunnova} proposed to begin a novel search for completely failed CCSNe by looking not for the presence but for the absence of sources. The ``Survey About Nothing'' monitors 1\E{6} red supergiants with the Large Binocular Telescope looking for the abrupt disappearance of any of these stars. In addition to potentially capturing a core-collapse failure, this survey could also detect VLE SNe coming from one of these sources. The supernovae themselves would be visible as a sudden brightening of the ``star" for of order a year, followed by a gradual but complete disappearance. After 4 years of observations \citet{gerke15} reviewed the survey data searching for both complete failures as well as the neutrino-mediated transients created by the Nadyozhin-Lovegrove effect. Four candidates were initially recovered, but followup observations ruled out three sources as they later reappeared. The final candidate event satisfied the criteria for a very low energy supernova and continued to be observed. Recently \citet{adams16} {\red reported 3 more years of observation on this candidate}, showing that the source had dimmed significantly below the progenitor luminosity. Modeling of possible dust effects compared to optical and IR source data suggest that the observed event was terminal and that the transient's cool, dim properties cannot be explained by dust. This event may therefore be the first observed example of the Nadyozhin-Lovegrove effect and an excellent example of a real VLE SNe. Unfortunately this survey's cadence is not frequent enough to catch a breakout event.

\citet{Rey15} conducted a search through HST archival data looking for collapse events that were not flagged by survey selection rules at the time and recovered one candidate in the range 25 - 35 \Msun that may have undergone an optically dark collapse.




\subsection{Candidate Shock Breakout Events}
\lSect{breakout_obs}

Several candidate shock breakout events have been published in the
literature, but most are high-energy events suspected to come from
compact progenitors. Soft X-ray breakout bursts are easier to
detect because of the large number of existing space-based X-ray
transient satellites designed specifically for the wide-field coverage
and rapid slew time needed to capture breakout. In 2008 \citet{Sod08}
serendipitously captured an X-ray transient when a supernova went off
during a Swift observation of its host galaxy. \citet{Sod08} attribute
this event to a Type Ib/c CCSN breaking out from a dense stellar wind
surrounding its progenitor, a scenario consistent with the high mass
loss rates of Type Ib/c progenitors near the end of their
lives. Unfortunately this rapid high-energy event, while of great
interest on its own and as a proof of concept for shock breakout
observations, bears little relation to the transients explored in this
work. 

Closer to the VLE SNe regime, UV observations using the GALEX
satellite in 2008 detected two CCSNe very close to the time of
explosion \citep{gezari08}: one with fading and one with rapidly
rising UV emission, suggesting that the latter had been caught during
its breakout phase. KEPLER hydrodynamic models combined with the
CMFGEN radiation transport code were used to model the observed UV
light curve of these events as breakout in a 15 \Msun\ red supergiant
exploding with a final kinetic energy 1.2 B. The calculated effective
temperature was similar to the high-energy RSG15 models presented
here. The authors noted an effect that will also be important in the
case of VLE SNe, namely that as the bolometric light curve fades the
spectral temperature declines, bringing more luminosity into the UV
(or optical) observing window. The net effect is an apparent plateau
phase that is actually the result of competing processes.

\subsubsection{Kepler Satellite Observations}

In early 2016 \citet{kepler2} announced the observation of two CCSNe
with the Kepler satellite\footnote{Not to be confused with the KEPLER
  stellar evolution code!} and provided data suggesting the telescope
had also captured the associated shock breakouts. The imaging cadence
of Kepler is still insufficient to resolve the breakout itself, but
subtracting models of the expected light curve from the data shows a
systematic excess consistent with a breakout event producing
additional luminosity at the beginning of the transient. While the
observed supernovae are much more energetic than the events modeled
here, the detections serve as an interesting proof of concept for
shock breakout observation, particularly for VLE SNe transients that
would have significantly longer timescales.

\subsection{Current \& Upcoming Observing Programs}
\lSect{observing}

The key to capturing these breakouts is high survey cadence, preferably hourly. Even a daily measurement can miss the more energetic breakouts entirely. Followup will be simpler than for compact star breakouts since observers will have a response window measured in hours rather than seconds. Spectroscopic followup is strongly recommended to measure \Tcol\!. The spectrum of a red supergiant breakout will be dominated by the blackbody continuum and hydrogen-helium lines. It may also show absorption features from the surrounding nebula. Unlike the later supernova light curve, the breakout will show little to no nickel or iron emission. The light curve will transition to photometric and spectroscopic behavior typical of a normal (albeit dim) Type II SN. Some planned missions could take good observations of breakout transients. The WFIRST mission would provide wide-field observations in the near-IR and the ULTRASAT program currently in design would launch a rapid-cadence UV satellite that would be ideal for detecting shock breakouts. Among ground-based observatories, Pan-STARRS, PTF/ZTF, LCOGT, and eventually LSST could all make useful observations, although none of these networks are a perfect fit.

\section{Conclusions}

\lSect{conclusions} 

Supernova shock breakout is a promising tool for detecting otherwise
dim supernova and retrieving information about their progenitor
stars. Observations - or null detections - of VLE SNe can help
understand the full range of CCSNe outcomes and place realistic,
observational constraints on the failure and partial failure rate of
supernovae. Shock breakouts in VLE SNe, in particular, may be easier
to observe than those in regular CCSNe, despite the larger bolometric
luminosity of the latter, because of their extended duration and
cooler spectral temperature. At the same time they are easier to
observe than their associated VLE SNe because of their higher
luminosity.


The shock breakout of SN1987A is modeled first as a test of the CASTRO
multigroup radiation transport module, giving peak luminosities,
durations, and color temperatures similar to other published
studies. Careful examination of the spectrum of this model motivates a
discussion of opacity, and of velocity terms in the radiation transport
equations. The impact of velocity on color temperature in low-opacity
high-energy models is illustrated.

Two red supergiants, RSG15 and RSG25, are selected for detailed study.
These masses approximately bracket the suspected mass range of failed
supernovae, though, in fact, it is the stellar radius and shock energy
that matter most. The choice of stellar atmosphere is considered and
found to have a potentially significant impact on breakout behavior. A
range of low-energy explosions is then modeled in KEPLER and
transferred to CASTRO. Two additional events with more standard energies
are also simulated in RSG15. All these events give light curves and
spectra that show clear variations with both explosion energy and
progenitor radius. Peak bolometric luminosities for VLE SNe range from
$10^{39} - 10^{44}$ erg/s depending on explosion energy and progenitor
mass. Peak values are compared to results from the KEPLER code and
from analytic predictions and found to be in reasonable agreement.

RSG15 models are then further simulated using CASTRO's multigroup
radiation transport module in order to determine the departure of
their spectrum from that calculated assuming radiation and matter are
in equilibrium at the photosphere. The relative roles of absorption
and scattering processes is found to be critical. Analytic formulas
for different absorption processes that become significant at the
lower temperatures common in VLE SNe are presented and
discussed. Analytic arguments suggest that \Tcol in VLE SNe will
behave differently from that in high-energy events. At lower energies,
the higher ratio of absorptive to scattering opacity processes will
cause \Tcol and \Teff to converge, resulting in a cooler spectrum than
would otherwise be predicted. {\red From a consideration of sources of
  absorptive opacity (Figs. 1 -3) and rough estimates of the
  thermalization criteria (\Sect{vle_tcol}), we estimate that breakout
  in RSG with kinetic energies less than about 10$^{49}$ erg will have
  nearly equal effective and color temperatures while those close to
  10$^{51}$ erg and above will have values like those determined with
  the neglect of absorption (\Tab{rsgresults}). Intermediate values of
  explosion energy will have color temperatures between $T_{ef}$ and
  \Tcol.  Further work using more realistic opacities is needed to
  give accurate quantitative results for these intermediate cases.}

VLE SNe breakout is expected to produce a blue ($>$ 1\E{4} K)
transient with an approximate duration 3 - 70 h and a bolometric
luminosity $10^{39} - 10^{44}$ erg. A slim but reasonable fraction of
this energy will be emitted at observable optical and IR wavelengths
because of the significantly lower color temperatures. Light curves
are presented for RSG15 models observed in IR, demonstrating that at
these wavelengths shock breakout in VLE SNe can in fact outshine its
more energetic counterparts. Current and upcoming observation programs
are assessed for suitability in detecting VLE SNe breakout. Cadence is
the limiting factor for both existing and future surveys; breakout
observations require cadence less than a day and preferably hourly.

\section{Acknowledgements}

We thank Ann Almgren and John Bell for developing the BOXLIB framework and the CASTRO code as well as Mike Zingale and Max Katz for their substantial contributions to it. Weiqun Zhang provided tireless and invaluable code support throughout this project. EL also thanks Jimmy Fung, Mack Kenamond, and Rob Lowrie for assistance in understanding flux-limited diffusion, and George Fuller for project support. Dan Kasen provided important discussions on the topic of opacity. Elizabeth Baumel provided debugging support; Katie Hamren pointed us towards the SVO Filter Profile Service, which provided clear, helpful, and thorough data on observing filters. This research has been supported by the NASA Theory Program (NNX14AH34G), the DOE High Energy Physics Program (DE-FC02-09ER41438) and a UC Lab Fees Research Award (12-LR-237070).

\bibliography{breakout}
\nocite{sandersII}
\end{document}